\documentclass[12pt,english]{article}
\usepackage[english]{babel}
\usepackage[utf8]{inputenc}
\usepackage{graphicx,xcolor}
\usepackage{epsf}
\usepackage{amsmath}
\usepackage{amssymb}
\usepackage{hyperref}
\usepackage{booktabs}
\usepackage{siunitx}
\usepackage{hyperref}
\usepackage{listings}
\newcommand{\beq}{\begin{equation}}
\newcommand{\eeq}{\end{equation}}
\newcommand{\bea}{\begin{eqnarray}}
\newcommand{\eea}{\end{eqnarray}}

\newcommand{\stringa}{\ttfamily\lstinline}
\def\cod#1{{\stringa!#1!}}

\begin{document}
\begin{titlepage}
\begin{center}
{\LARGE \bf Mueller-Navelet jets at 13 TeV LHC:
dependence on dynamic constraints in the central rapidity region}
\end{center}

\vskip 0.5cm

\centerline{F.G.~Celiberto$^{1,2\ast}$, D.Yu.~Ivanov$^{3,4\P}$, B.~Murdaca$^{2\dag}$
and A.~Papa$^{1,2\ddagger}$}

\vskip .6cm

\centerline{${}^1$ {\sl Dipartimento di Fisica, Universit\`a della Calabria,}}
\centerline{\sl Arcavacata di Rende, I-87036 Cosenza, Italy}

\vskip .2cm

\centerline{${}^2$ \sl Istituto Nazionale di Fisica Nucleare, Gruppo collegato 
di Cosenza,}
\centerline{\sl Arcavacata di Rende, I-87036 Cosenza, Italy}

\vskip .2cm

\centerline{${}^3$ \sl Sobolev Institute of Mathematics, 630090 Novosibirsk, 
Russia}

\vskip .2cm

\centerline{${}^4$ \sl Novosibirsk State University, 630090 Novosibirsk, Russia}

\vskip 2cm

\begin{abstract}
We study the production of Mueller-Navelet jets at 13 TeV LHC, within collinear
factorization and including the BFKL resummation of energy logarithms
in the next-to-leading approximation.
We calculate several azimuthal correlations for different values of the
rapidity separation $Y$ between the two jets and evaluate the effect of
excluding those events where, for a given $Y$, one of the two jets is produced
in the central region.
\end{abstract}

%\vskip .5cm

$
\begin{array}{ll}
^{\ast}\mbox{{\it e-mail address:}} &
\mbox{francescogiovanni.celiberto@fis.unical.it}\\
^{\P}\mbox{{\it e-mail address:}} &
\mbox{d-ivanov@math.nsc.ru}\\
^{\dag}\mbox{{\it e-mail address:}} &
\mbox{beatrice.murdaca@cs.infn.it}\\
^{\ddagger}\mbox{{\it e-mail address:}} &
\mbox{alessandro.papa@fis.unical.it}\\
\end{array}
$

\end{titlepage}

\vfill \eject

\section{Introduction}
\label{intro}

The production at the LHC of Mueller-Navelet jets~\cite{Mueller:1986ey} 
represents a fundamental test of QCD at high energies.
It is an inclusive process where two jets, characterized by large transverse 
momenta that are of the same order and much larger than $\Lambda_{\rm QCD}$, 
are produced in proton-proton collisions, separated by a large rapidity gap 
$Y$ and in association with an undetected hadronic system $X$.

At the LHC energies the rapidity gap between the two jets can be large enough,
that the emission of several undetected hard partons, having large transverse 
momenta, with rapidities intermediate to those of the two detected jets, 
becomes possible. The probability of
this emission is suppressed in perturbation theory by one power of $\alpha_s$
per produced parton, but when final-state partons are strongly ordered
in rapidity, it is also enhanced by large logarithms of the energy which
can compensate the smallness of the QCD coupling.

The BFKL approach~\cite{BFKL} provides with a systematic framework
for the resummation of these energy logarithms, both in the leading
logarithmic approximation (LLA), which means all terms
$(\alpha_s\ln(s))^n$, and in the next-to-leading logarithmic approximation
(NLA), which means resummation of all terms $\alpha_s(\alpha_s\ln(s))^n$.
In this approach, the cross section for Mueller-Navelet jet production takes
the form of a convolution between two impact factors for the transition from
each colliding proton to the forward jet (the so-called ``jet vertices'') and
a process-independent Green's function.

The BFKL Green's function obeys an iterative integral equation, whose
kernel is known at the next-to-leading order (NLO) both for forward
scattering ({\it i.e.} for $t=0$ and color singlet in the
$t$-channel)~\cite{FL98,CC98} and for any fixed (not growing with energy)
momentum transfer $t$ and any possible two-gluon color state in the
$t$-channel~\cite{Fadin:1998jv,FG00,FF05}.

\begin{figure}[tb]
\centering
\includegraphics[scale=0.7]{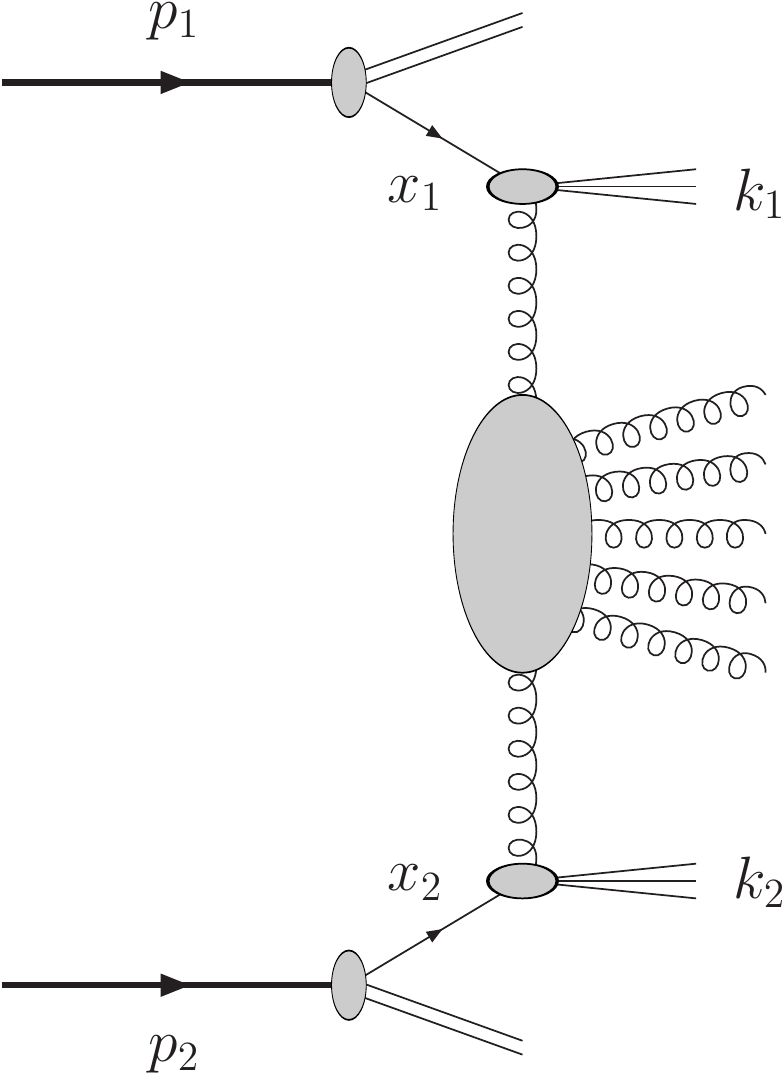}
\caption[]{%Schematic representation of the
Mueller-Navelet jet production process.}
\label{fig:MN}
\end{figure}

The jet vertex can be expressed, within collinear factorization at the
leading twist, as the convolution of parton distribution functions (PDFs)
of the colliding proton, obeying the standard DGLAP evolution~\cite{DGLAP},
with the hard process describing the transition from the parton emitted
by the proton to the forward jet in the final state. The Mueller-Navelet
jet production process is, therefore, a unique venue, where the two main
resummation mechanisms of perturbative QCD play their role at the same time
(see Fig.~\ref{fig:MN} for a schematic view).

The expression for the ``jet vertices'' was first obtained with NLO accuracy
in~\cite{Bartels:2002yj}, a result later confirmed in~\cite{Caporale:2011cc}.
A simpler expression, more practical for numerical purposes, was
obtained in~\cite{Ivanov:2012ms} within the so-called ``small-cone''
approximation (SCA)~\cite{Furman:1981kf,Aversa}, {\it i.e.} for small jet cone
aperture in the rapidity-azimuthal angle plane. The implementation of
several jet reconstruction algorithms, both in the exact jet vertex and in its
``small-cone'' version, has been carried out in~\cite{Colferai:2015zfa}.

A lot of papers have appeared, so far, about the Mueller-Navelet jet production
process at LHC, both at a center-of-mass energy of
14~TeV~\cite{Colferai2010,Caporale2013,Salas2013} and
7~TeV~\cite{Ducloue2013,Ducloue2014,Ducloue:2014koa,Caporale:2014gpa,Celiberto:2015yba}. Their main aim was the study of the $Y$-dependence of azimuthal angle
correlations between the two measured jets, {\it i.e.} average values of
$\cos{(n \phi)}$, where $n$ is an integer and $\phi$ is the angle in the
azimuthal plane between the direction of one jet and the opposite direction
of the other jet, and also of ratios of two such cosines~\cite{sabioV}.
These studies share the approach and the factorized form of the basic
amplitude, but differ in the setup of the jet vertex (exact or small-cone
approximated) and/or in the procedure to optimize the highly-unstable
BFKL series~\footnote{It is worth mentioning two recent studies,
Refs.~\cite{Ducloue:2015jba,Mueller:2015ael}, which considered, respectively,
the contribution to Mueller-Navelet jet production from the
double-parton exchange mechanism and from Sudakov resummations.}.
Several possibilities were considered: (i) the inclusion of pieces
of the (unknown) next-to-NLO corrections, as dictated by
\emph{collinear improvement}~\cite{collinear} or by energy-momentum
conservation~\cite{Kwiecinski:1999yx}, (ii) a suitable choice,
within the NLA accuracy, of the renormalization and factorization scales,
$\mu_R$ and $\mu_F$, and of the BFKL energy scale, $s_0$, some common
options being those inspired by the \emph{principle of minimum
sensitivity} (PMS)~\cite{PMS}, the \emph{fast apparent convergence}
(FAC)~\cite{FAC} and the \emph{Brodsky-LePage-Mackenzie method}
(BLM)~\cite{BLM} (see also Ref.~\cite{Caporale:2015uva}).
There is a clear evidence that theoretical results can nicely reproduce CMS
data~\cite{CMS} at 7~TeV in the range $5 \lesssim Y \lesssim 9.4$ when the
BLM optimization method is adopted, both in the implementation of the
amplitude with the exact jet vertex and collinearly-improved BFKL Green's
function (see, {\it e.g.}, Ref.~\cite{Ducloue2014}) and with the
small-cone jet vertex and no collinear improvement (see
Refs.~\cite{Caporale:2014gpa,Caporale:2015uva}), though the experimental
uncertainties on the azimuthal correlations and on the PDFs do not allow
to rule out the other optimization procedures. An important clarification
could come from the CMS analyses at 13~TeV and 14~TeV, since the larger
available energy in the center of mass implies the possibility of a larger
average number of parton emission between the jets and, hence, better
conditions for the manifestation of the BFKL dynamics. Moreover, some
added information could come (i) from the measurement, in addition to azimuthal
correlations, of the total cross section for Mueller-Navelet jets and (ii) from
the consideration of asymmetric cuts in the transverse momenta of the two
detected jets. It was indeed shown in Ref.~\cite{Caporale:2014gpa} that
the total cross section is much more sensitive to the optimization procedure
than azimuthal correlations and is, therefore, a better discriminator of
the various options. It was also discussed that the use of asymmetric cuts
in jet transverse momenta allows a better separation between
BFKL-resummed and fixed-order predictions in azimuthal correlations and their
ratios, as it was indeed shown in Ref.~\cite{Celiberto:2015yba}.

There is another issue which deserves some care and has not been taken
into consideration both in theoretical and experimental analyses so far.
As discussed in the last Section of Ref.~\cite{Caporale:2014gpa},
in defining the $Y$ value for a given final state with two jets, the rapidity
of one of the two jets could be so small, say $|y_i|\lesssim 2$, that this jet
is actually produced in the central region, rather than in one of the two
forward regions. Since the longitudinal momentum fractions of the parent
partons $x$ that generate such central jet are very small, one can naturally 
expect sizable corrections to the vertex of this jet, due to the fact that the
collinear factorization approach used in the derivation of the result for jet
vertex could not be accurate enough in our kinematic region, where $x$ values 
can be as small as, $x \sim 10^{-3}$.

The use of collinear factorization methods in the case of central jet production
in our kinematic range deserves some discussion. On one hand, at 
$x \sim 10^{-3}$ and at scales of the order of the jet transverse momenta which 
we consider here, $\sim 20\div 40$~GeV, PDFs are well constrained, mainly from 
DIS HERA data. On the other hand, in this kinematic region PDF 
parametrizations extracted in NNLO and in NLO approximations start to differ 
one from the other, which indicates that NNLO effects become essential in the 
DIS cross sections. The situation with central jet production in proton-proton 
collisions may be different. Recently in~\cite{Currie:2013dwa} results 
for NNLO corrections to the dijet production originated from the gluonic 
subprocesses were presented. In the region $|y_{1,2}|<0.3$ and for jet 
transverse momenta $\sim 100$~GeV, the account of NNLO effects leads to an
increase of the cross section by $\sim 25\%$. For our kinematics, 
featuring smaller jet transverse momenta and ``less inclusive'' coverage of jet 
rapidities, one could expect even larger NNLO corrections.               

Conceptually, instead of the collinear approach, for jets produced in the 
central rapidity region (at very small $x$) a promising approach would be to 
use a high-energy factorization scheme (often also referred as 
$k_T$-factorization) together with the NLO central jet vertex calculated 
in~\cite{Bartels:2006hg}~\footnote{For the discussion of different approaches 
to factorization for dijet production see, {\it e.g.}, the recent review 
paper~\cite{Sapeta:2015gee}.}.

Returning back to our case of Mueller-Navelet jets, we see here as an important 
task to reveal dynamic mechanisms for the partonic interaction in the 
semihard region, $s\gg |t|$, comparing theory predictions with data. From the 
theory side we have now the BFKL approach, where one can resum in a 
model-independent way only the leading and first subleading logarithms of the
energy. Several approaches to handle big effects beyond the NLA BFKL were 
suggested, such as the above-discussed collinear improvement, BLM, and so on. 
The comparison of theory predictions with experiment should clarify what is
the better approach. For this reason we suggest to compare BFKL theory 
predictions with data in a region where theoretical uncertainties related with 
other kind of physics are most possibly reduced. Therefore we propose to 
return to the original Mueller-Navelet idea, to study the inclusive production 
of two forward jets separated by a large rapidity gap, and to remove from the 
analysis those regions where jets are produced at central rapidities.   

As a contribution to the assessment of this effect, in this paper we will study 
the $Y$-dependence of several azimuthal correlations and ratios among them, 
imposing an additional constraint, that the rapidity of a Mueller-Navelet jet 
cannot be smaller than a given value. Then we will compare this option with 
the case when the constraint is absent. 

Since here we want to focus just on the possible impact of jets produced in the
central region, we will stick to a definite optimization setup, namely the
BLM one, which performed quite successfully in the comparison with CMS
data at 7~TeV. We will implement its ``exact'' version, according to the
nomenclature introduced in Ref.~\cite{Caporale:2015uva} and fix the
center-of-mass energy at 13~TeV, so that our results can be directly
compared with the forthcoming CMS analyses.

The paper is organized as follows: in the next section we recall the
kinematics and the basic formulae for the Mueller-Navelet jet process
cross section; in section~\ref{results} we present our results;
finally, in section~\ref{conclusions} we draw our conclusions.

\section{Theoretical setup}
\label{theory}

In this section we briefly recall the kinematics of the process and the
main formulae, referring the reader to previous
papers~\cite{Caporale2013,Caporale:2014gpa} for the omitted details.

The process under exam is the production of Mueller-Navelet
jets~\cite{Mueller:1986ey} in proton-proton collisions
\begin{eqnarray}
\label{process}
p(p_1) + p(p_2) \to {\rm jet}(k_{J_1}) + {\rm jet}(k_{J_2})+ X \;,
\end{eqnarray}
where the two jets are characterized by high transverse momenta,
$\vec k_{J_1}^2\sim \vec k_{J_2}^2\gg \Lambda_{\rm QCD}^2$ and large separation
in rapidity; $p_1$ and $p_2$ are taken as Sudakov vectors satisfying
$p_1^2=p_2^2=0$ and $2\left( p_1 p_2\right)=s$, working at leading twist and 
neglecting the proton mass and other power suppressed corrections.

In QCD collinear factorization the cross section of the process~(\ref{process})
reads
\beq
\frac{d\sigma}{dx_{J_1}dx_{J_2}d^2k_{J_1}d^2k_{J_2}}
=\sum_{i,j=q,{\bar q},g}\int_0^1 dx_1 \int_0^1 dx_2\ f_i\left(x_1,\mu_F\right)
\ f_j\left(x_2,\mu_F\right)\frac{d{\hat\sigma}_{i,j}\left(x_1x_2s,\mu_F\right)}
{dx_{J_1}dx_{J_2}d^2k_{J_1}d^2k_{J_2}}\;,
\eeq
where the $i, j$ indices specify the parton types (quarks $q = u, d, s, c, b$;
antiquarks $\bar q = \bar u, \bar d, \bar s, \bar c, \bar b$; or gluon $g$),
$f_i\left(x, \mu_F \right)$ denotes the initial proton PDFs; $x_{1,2}$ are
the longitudinal fractions of the partons involved in the hard subprocess,
while $x_{J_{1,2}}$ are the jet longitudinal fractions; $\mu_F$ is the
factorization scale; $d\hat\sigma_{i,j}\left(x_1x_2s, \mu_F \right)$ is
the partonic cross section for the production of jets and
$x_1x_2s\equiv\hat s$ is the squared center-of-mass energy of the
parton-parton collision subprocess (see Fig.~\ref{fig:MN}).

The cross section of the process can be presented as
\beq
\frac{d\sigma}
{dy_{J_1}dy_{J_2}\, d|\vec k_{J_1}| \, d|\vec k_{J_2}|
d\phi_{J_1} d\phi_{J_2}}
=\frac{1}{(2\pi)^2}\left[{\cal C}_0+\sum_{n=1}^\infty  2\cos (n\phi )\,
{\cal C}_n\right]\, ,
\eeq
where $\phi=\phi_{J_1}-\phi_{J_2}-\pi$, while ${\cal C}_0$ gives the total
cross section and the other coefficients ${\cal C}_n$ determine the distribution
of the azimuthal angle of the two jets.

Since the main object of the present analysis is the impact of jet produced
in the central region on azimuthal coefficients, we will adopt 
just one representation for ${\cal C}_n$, out of the many possible
NLA-equivalent options (see Ref.~\cite{Caporale:2014gpa} for a discussion).
In particular, we will use the so-called {\em exponentiated} representation
together with the BLM optimization method to fix the common value for the 
renormalization scale $\mu_R$ and the factorization scale $\mu_F$. 
In~\cite{Caporale:2014gpa} it was shown that
this setup allows a nice agreement with CMS data for several azimuthal
correlations and their ratios in the large $Y$ regime.
In our calculation we will use ``exact''   and in some cases
also approximate, semianalytic   implementations of BLM method,
which called below as $(a)$, $(b)$ cases, in order to keep contact with with 
previous applications of BLM method where approximate approaches were used, 
for the details see Ref.~\cite{Caporale:2015uva}.

Introducing, for the sake of brevity, the definitions
\[
Y=y_1-y_2=\ln\frac{x_{J_1}x_{J_2}s}{|\vec k_{J_1}||\vec k_{J_2}|}\;,
\;\;\;\;\;
Y_0=\ln\frac{s_0}{|\vec k_{J_1}||\vec k_{J_2}|}\;,
\]
we will present in what follows 
the three different expressions for the coefficients ${\cal C}_n$.

$\bullet$ case ``exact'' 

The BLM optimal scale $\mu_R^{\rm BLM}$ is defined as the value of $\mu_R$ that 
makes all contributions to the considered observables which are  
proportional to the QCD $\beta-$ function, $\beta_0$, vanish. In our case we 
have 
\[
{\cal C}^{\beta}_n
\equiv \frac{x_{J_1}x_{J_2}}{|\vec k_{J_1}||\vec k_{J_2}|}
\int\limits^{\infty}_{-\infty} d\nu
\left(\frac{s}{s_0}\right)^{\bar \alpha^{\rm MOM}_s(\mu^{\rm BLM}_R)\chi(n,\nu)}
\left(\alpha^{\rm MOM}_s (\mu^{\rm BLM}_R)\right)^3
\]
\beq\label{c_nnnbeta}
\times c_1(n,\nu)c_2(n,\nu) \frac{\beta_0}{2 N_c} \left[\frac{5}{3}
+\ln \frac{(\mu^{\rm BLM}_R)^2}{Q_1 Q_2} -2\left( 1+\frac{2}{3}I \right)
\right.
\eeq
\[
\left.
+\bar \alpha^{\rm MOM}_s(\mu^{\rm BLM}_R)\ln\frac{s}{s_0} \: \frac{\chi(n,\nu)}{2}
\left(-\frac{\chi(n,\nu)}{2}+\frac{5}{3}+\ln \frac{(\mu^{\rm BLM}_R)^2}{Q_1 Q_2}
-2\left( 1+\frac{2}{3}I \right)\right)\right]=0 \, .
\]
The first term in the r.h.s. of Eq.~(\ref{c_nnnbeta}) originates from the NLO 
correction to the jet vertices, whereas the second, $\sim  \alpha^{\rm MOM}$, 
contribution is due to the $\sim \beta_0$ part of NLO correction to the kernel 
of the BFKL equation.

In~\cite{Caporale:2015uva} we considered the implementation of the BLM method 
for general semihard process. We found that the above-mentioned 
$\sim \beta_0$-contributions to the NLO impact factors are universally 
expressed in terms of the LO impact factors of the considered process 
(the LO jet vertices for the Mueller-Navelet process considered here). 
Such contributions must be taken into account in the implementation of 
BLM method, because {\it all} contributions to the cross section that 
are $\sim \beta_0$ must vanish at the BLM scale.  

After that we have the following expression for our observables:
\[
{\cal C}_n^{\rm BLM}=\frac{x_{J_1}x_{J_2}}{|\vec k_{J_1}||\vec k_{J_2}|}
\int_{-\infty}^{+\infty}d\nu \ e^{(Y-Y_0) \bar \alpha_s^{\rm MOM}(\mu_R^{\rm BLM})
\left[\chi\left(n,\nu\right)+\bar\alpha_s^{\rm MOM}(\mu_R^{\rm BLM})
\left(\bar\chi\left(n,\nu\right)+\frac{T^{\rm conf}}{N_c}\chi(n,\nu)\right)
\right]}
\]
\beq\label{blm_exact}
\times (\alpha_s^{\rm MOM}(\mu_R^{\rm BLM}))^2
c_1(n,\nu,|\vec k_{J_1}|, x_{J_1}) c_2(n,\nu,|\vec k_{J_2}|,x_{J_2})
\eeq
\[
\times \left[1 + \alpha_s^{\rm MOM}(\mu_R^{\rm BLM})
\left\{\frac{\bar c_1^{\left(1\right)}(n,\nu,|\vec k_{J_1}|,x_{J_1})}
{c_1(n,\nu,|\vec k_{J_1}|, x_{J_1})}
+\frac{\bar c_2^{\left(1\right)}(n,\nu,|\vec k_{J_2}|, x_{J_2})}
{c_2(n,\nu,|\vec k_{J_2}|, x_{J_2})}+\frac{2T^{\rm conf}}{N_c}\right\}\right]\;,
\]
In the above equations, $\bar \alpha_s^{\rm MOM}\equiv \alpha_s^{\rm MOM} N_c/\pi$,
with $N_c$ the number of colors and $\alpha_s^{\rm MOM}$ is the QCD coupling 
in the physical momentum subtraction (MOM) scheme, related to 
$\alpha_s^{\overline{\rm MS}}$ by a finite renormalization,
\beq
\alpha_s^{\overline{\rm MS}}=\alpha_s^{\rm MOM}\left(1+\frac{\alpha_s^{\rm MOM}}{\pi}T
\right)\;,
\eeq
with $T=T^{\beta}+T^{\rm conf}$,
\beq
T^{\beta}=-\frac{\beta_0}{2}\left( 1+\frac{2}{3}I \right)\, ,
\eeq
\[
T^{\rm conf}= \frac{N_c}{8}\left[ \frac{17}{2}I +\frac{3}{2}\left(I-1\right)\xi
+\left( 1-\frac{1}{3}I\right)\xi^2-\frac{1}{6}\xi^3 \right] \;,
\]
where $I=-2\int_0^1dx\frac{\ln\left(x\right)}{x^2-x+1}\simeq2.3439$ and $\xi$
is a gauge parameter, fixed at zero in the following. Then,
\beq
\beta_0=\frac{11}{3} N_c - \frac{2}{3}n_f
\eeq
is the first coefficient of the QCD $\beta$-function,
\beq
\chi\left(n,\nu\right)=2\psi\left( 1\right)-\psi\left(\frac{n}{2}
+\frac{1}{2}+i\nu \right)-\psi\left(\frac{n}{2}+\frac{1}{2}-i\nu \right)
\eeq
is the LO BFKL characteristic function,
\beq
\label{c1}
c_1(n,\nu,|\vec k|,x)=2\sqrt{\frac{C_F}{C_A}}
(\vec k^{\,2})^{i\nu-1/2}\,\left(\frac{C_A}{C_F}f_g(x,\mu_F)
+\sum_{a=q,\bar q}f_a(x,\mu_F)\right)
\eeq
and
\beq
\label{c2}
c_2(n,\nu,|\vec k|,x)=\biggl[c_1(n,\nu,|\vec k|,x) \biggr]^* \;,
\eeq
are the LO jet vertices in the $\nu$-representation. The remaining objects
are related with the NLO corrections of the BFKL kernel ($\bar \chi(n,\nu)$,
given in Eqs.~(23) of Ref.~\cite{Caporale2013}) and of the jet vertices in
the small-cone approximation ($c_{1,2}^{(1)}(n,\nu,|\vec k_{J_{1,2}}|, x_{J_{1,2}})$,
given in Eqs.~(36) and~(37) of Ref.~\cite{Caporale2013}).
The functions $\bar c_{1,2}^{(1)}(n,\nu,|\vec k_{J_2}|, x_{J_2})$ are the same as
$c_{1,2}^{(1)}(n,\nu,|\vec k_{J_{1,2}}|, x_{J_{1,2}})$ with all terms proportional to
$\beta_0$ removed.

Note that, the ``exact'' implementation of the BLM method requires numerical  
solution of an integral equation,  Eq.~(\ref{c_nnnbeta}) for each value of $s$
and the values of $\mu_R^{\rm BLM}$ obtained in this way depend on the energy 
of the process.

Below we will perform also calculations with two approximated approaches to the 
BLM scale setting. We will consider the options where $\mu_R$ is chosen such 
that in the r.h.s. of Eq.~(\ref{c_nnnbeta}) either the term coming from the NLO 
correction to the jet vertices vanishes [case $(a)$], or the contribution 
due to the $\sim \beta_0$ part of NLO BFKL kernel does [case $(b)$].  
In these two cases one gets simpler analytical expressions for the BLM scales 
which do not depend on the energy. Such approximate approaches were used 
earlier in the literature of the BLM method for different semihard processes 
(see a more detailed discussion in~\cite{Caporale:2015uva}). Here we will 
perform also some calculations with these approximate schemes (a) and (b), 
in order to get an idea about the inaccuracy of the predictions for 
Mueller-Navelet jets observables related with such approximate 
implementations of the BLM scale setting.  

So, we have: 

$\bullet$ case $(a)$
\[
(\mu_{R,a}^{\rm BLM})^2 = k_{J_1} k_{J_2} \exp\left[2\left(1+\frac{2}{3}I\right)
-\frac{5}{3}\right]\;,
\]
with 
\[
{\cal C}_n^{\rm BLM, a}= \frac{x_{J_1}x_{J_2}}{|\vec k_{J_1}||\vec k_{J_2}|}
\int_{-\infty}^{+\infty}d\nu \ e^{(Y-Y_0)
\left[\bar \alpha_s^{\rm MOM}(\mu_{R,a}^{\rm BLM})\chi(n,\nu)
+ (\bar{\alpha}_s^{\rm MOM}(\mu_{R,a}^{\rm BLM}))^2
\left( \bar \chi(n,\nu)+\frac{T^{\rm conf}}{N_c}\chi(n,\nu)
- \frac{\beta_0}{8N_c}\chi^2(n,\nu)\right)\right]}
\]
\beq\label{casea}
\times (\alpha_s^{\rm MOM}(\mu_{R,a}^{\rm BLM}))^2
c_1(n,\nu,|\vec k_{J_1}|, x_{J_1}) c_2(n,\nu,|\vec k_{J_2}|,x_{J_2})
\eeq
\[
\times \left[1 + \alpha_s^{\rm MOM}(\mu_{R,a}^{\rm BLM})
\left\{\frac{\bar c_1^{\left(1\right)}(n,\nu,|\vec k_{J_1}|,x_{J_1})}
{c_1(n,\nu,|\vec k_{J_1}|, x_{J_1})}
+\frac{\bar c_2^{\left(1\right)}(n,\nu,|\vec k_{J_2}|, x_{J_2})}
{c_2(n,\nu,|\vec k_{J_2}|, x_{J_2})}+\frac{2T^{\rm conf}}{N_c}\right\}\right]\;,
\]

and

$\bullet$ case $(b)$ 
\[
(\mu_{R,b}^{\rm BLM})^2 = k_{J_1} k_{J_2} \exp\left[2\left(1+\frac{2}{3}I\right)
-\frac{5}{3}+\frac{1}{2}\, \chi(n,\nu)\right]\;,
\]
with
\[
{\cal C}_n^{\rm BLM, b}= \frac{x_{J_1}x_{J_2}}{|\vec k_{J_1}||\vec k_{J_2}|}
\int_{-\infty}^{+\infty}d\nu \ e^{(Y-Y_0)
\left[\bar \alpha_s^{\rm MOM}(\mu_{R,b}^{\rm BLM})\chi(n,\nu)
+ (\bar{\alpha}_s^{\rm MOM}(\mu_{R,b}^{\rm BLM}))^2
\left( \bar \chi(n,\nu)+\frac{T^{\rm conf}}{N_c}\chi(n,\nu)\right)\right]}
\]
\beq\label{caseb}
\times (\alpha_s^{\rm MOM}(\mu_{R,b}^{\rm BLM}))^2
c_1(n,\nu,|\vec k_{J_1}|, x_{J_1}) c_2(n,\nu,|\vec k_{J_2}|,x_{J_2})
\eeq
\[
\times \left[1 + \alpha_s^{\rm MOM}(\mu_{R,b}^{\rm BLM})
\left\{\frac{\bar c_1^{\left(1\right)}(n,\nu,|\vec k_{J_1}|,x_{J_1})}
{c_1(n,\nu,|\vec k_{J_1}|, x_{J_1})}
+\frac{\bar c_2^{\left(1\right)}(n,\nu,|\vec k_{J_2}|, x_{J_2})}
{c_2(n,\nu,|\vec k_{J_2}|, x_{J_2})}+\frac{2T^{\rm conf}}{N_c}
+ \frac{\beta_0}{4N_c}\chi\left(n,\nu\right) \right\}\right]\;.
\]

Note that, in the above equations the scale $s_0$ entering $Y_0$ is the 
artificial energy scale introduced in the BFKL approach to perform the Mellin 
transform from the $s$-space to the complex angular momentum plane and cancels 
in the full expression, up to terms beyond the NLA. In the following it will 
always be fixed at the ``natural'' value $Y_0=0$, given by the kinematic of 
Mueller-Navelet process.

\section{Numerical analysis}
\label{results}

In this Section we present our results for the dependence on the
rapidity separation between the detected jets, $Y=y_{J_1}-y_{J_2}$,
of ratios ${\cal R}_{nm}\equiv{\cal C}_n/{\cal C}_m$ between the
coefficients ${\cal C}_n$. Among them, the ratios of
the form $R_{n0}$ have a simple physical interpretation, being the azimuthal
correlations $\langle \cos(n\phi)\rangle$.

In order to match the kinematic cuts used by the CMS collaboration, we will
consider the \emph{integrated coefficients} given by
\begin{equation}\label{Cm_int}
\begin{aligned}
&
C_n=\int_{y_{1,\rm min}}^{y_{1,\rm max}}dy_1
\int_{y_{2,\rm min}}^{y_{2,\rm max}}dy_2\int_{k_{J_1,\rm min}}^{\infty}dk_{J_1}
\int_{k_{J_2,\rm min}}^{\infty}dk_{J_2}
\\ &
\delta\left(y_1-y_2-Y\right)
\theta\left(|y_1| - y^{\rm C}_{\rm max}\right) \theta\left(|y_2| - y^{\rm C}_{\rm max}
\right) {\cal C}_n \left(y_{J_1},y_{J_2},k_{J_1},k_{J_2} \right)
\end{aligned}
\end{equation}
and their ratios $R_{nm}\equiv C_n/C_m$. In Eq.~(\ref{Cm_int}),
the two step-functions force the exclusion of jets whose rapidity is smaller
than a cutoff value, given by $y^{\rm C}_{\rm max}$, which delimits the central
rapidity region. We will take jet rapidities in the range delimited by
$y_{1,\rm min}=y_{2,\rm min}=-4.7$  and $y_{1,\rm max}=y_{2,\rm max}=4.7$, as in the
CMS analyses at 7~TeV, and consider $Y=3.5$, 4.5, 5.5, 6.5, 7.5, 8.5, 9.0.

As for the values of $y^{\rm C}_{\rm max}$, we will consider three cases:
$y^{\rm C}_{\rm max}=0$, which means no exclusion from jets in the central
region, as in all the numerical analyses so far; $y^{\rm C}_{\rm max}=1.5$,
corresponding to a central region with size equal to about one third of the
maximum possible rapidity span $Y=9.4$ and $y^{\rm C}_{\rm max}=2.5$, as a
control value, to check the stability of our results.

Concerning the jet transverse momenta, differently from most previous analyses,
we make the following five choices, which include \emph{asymmetric} cuts:
(1) $k_{J_1,\rm min} = 20$ GeV, $k_{J_2,\rm min} = 20$ GeV,
(2) $k_{J_1,\rm min} = 20$ GeV, $k_{J_2,\rm min} = 30$ GeV,
(3) $k_{J_1,\rm min} = 20$ GeV, $k_{J_2,\rm min} = 35$ GeV,
(4) $k_{J_1,\rm min} = 20$ GeV, $k_{J_2,\rm min} = 40$ GeV, and
(5) $k_{J_1,\rm min} = 35$ GeV, $k_{J_2,\rm min} = 35$ GeV.
The jet cone size $R$ entering the NLO-jet vertices is fixed at the value
$R=0.5$, the center-of-mass energy at $\sqrt s=13$ TeV and, as anticipated,
$Y_0=0$. We use the PDF set MSTW 2008 NLO~\cite{PDF} and the two-loop
running coupling with $\alpha_s\left(M_Z\right)=0.11707$. 
The MSTW 2008 NLO PDF set was used successfully in various analyses
of inclusive jet production at LHC, including our previous studies of 
Mueller-Navelet jets. Now there exist updated PDF parametrizations, including 
the MMHT 2014 set~\cite{PDF2}, which is the successor of the MSTW 2008 
analysis. Here we continue to use MSTW 2008 NLO PDFs
because in our kinematic range the difference between MSTW 2008 NLO and the 
updated MMHT 2014 NLO PDFs is very small. Also, we want to keep the 
opportunity to compare our results at 13~TeV with our previous calculations at 
7~TeV without introducing any other source of discrepancy related to the change
of the PDF set.

All numerical calculations were implemented in \textsc{Fortran}.
Numerical integrations and the computation of the polygamma functions
were performed using specific \textsc{CERN} program libraries~\cite{cernlib}.
Furthermore, we used slightly modified versions of the \cod{Chyp}~\cite{chyp}
and \cod{Psi}~\cite{rpsi} routines in order to perform the calculation
of the Gauss hypergeometric function $_2F_1 $ and of the real part of the
$\psi$ function, respectively.

The most significant source of uncertainty is the numerical 4-dimensional
integration over the variables
$|\vec k_{J_1}|$, $|\vec k_{J_2}|$, $y_{J_1}$ and
$\nu$, which was directly estimated by \cod{Dadmul} integration
routine~\cite{cernlib}.
In a recent paper~\cite{Celiberto:2015yba}, we have shown that
the other two sources, which are respectively
the one-dimensional integration over the longitudinal momentum fraction
$\zeta$ in the NLO impact factors $c_{1,2}^{(1)}(n,\nu,|\vec k_{J_{1,2}}|,
x_{J_{1,2}})$ (see Eqs.~(36) and~(37) of Ref.~\cite{Caporale2013})
and the upper cutoff in the numerical integrations over
$|\vec k_{J_1}|$, $|\vec k_{J_2}|$ and $\nu$, are negligible with respect to
the first one. For this reason the error bars of all predictions presented 
in this work are just those given by the \cod{Dadmul} routine.

\begin{figure}[p]
\centering

   \includegraphics[scale=0.40]{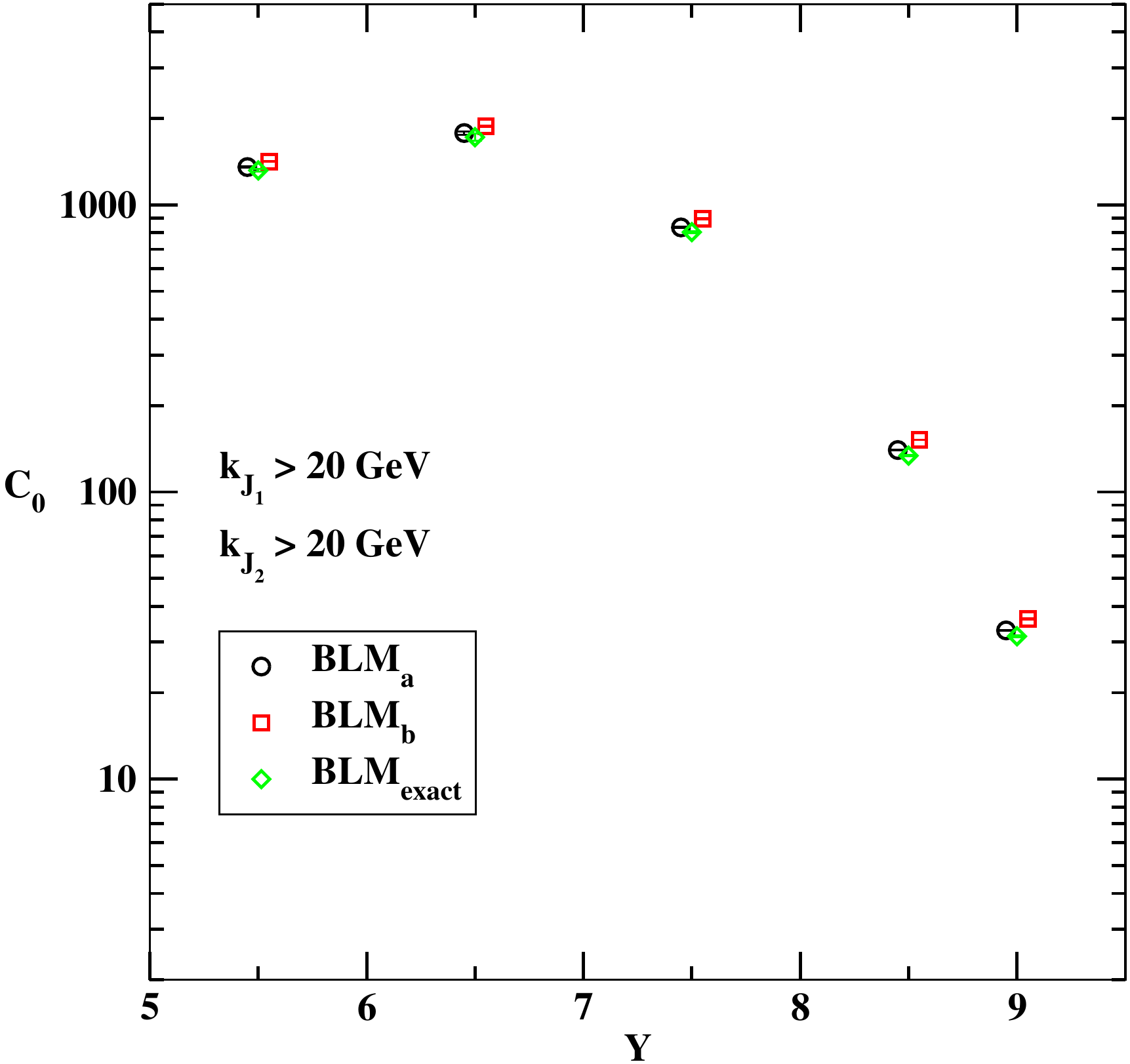}
   \includegraphics[scale=0.40]{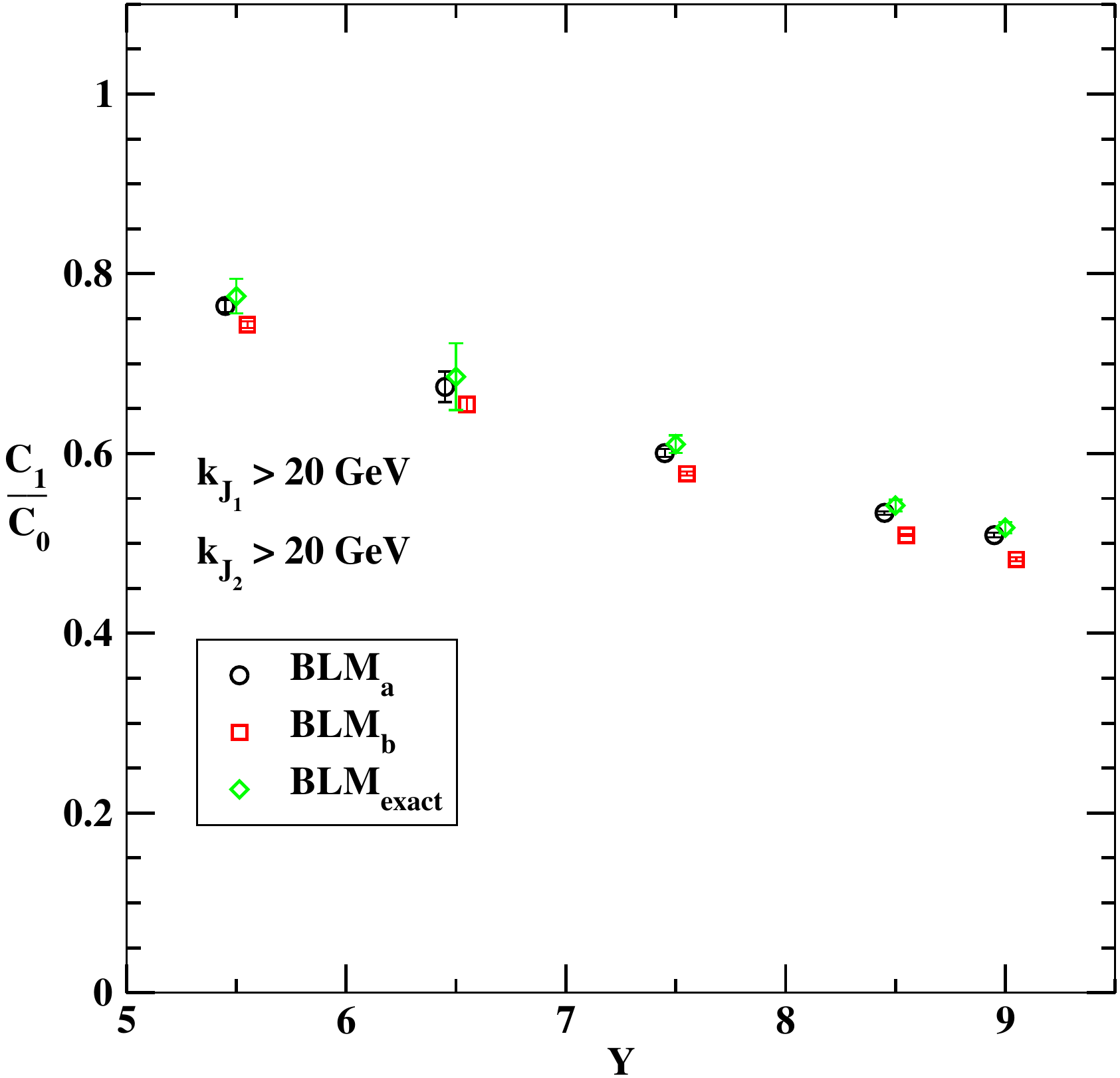}

   \includegraphics[scale=0.40]{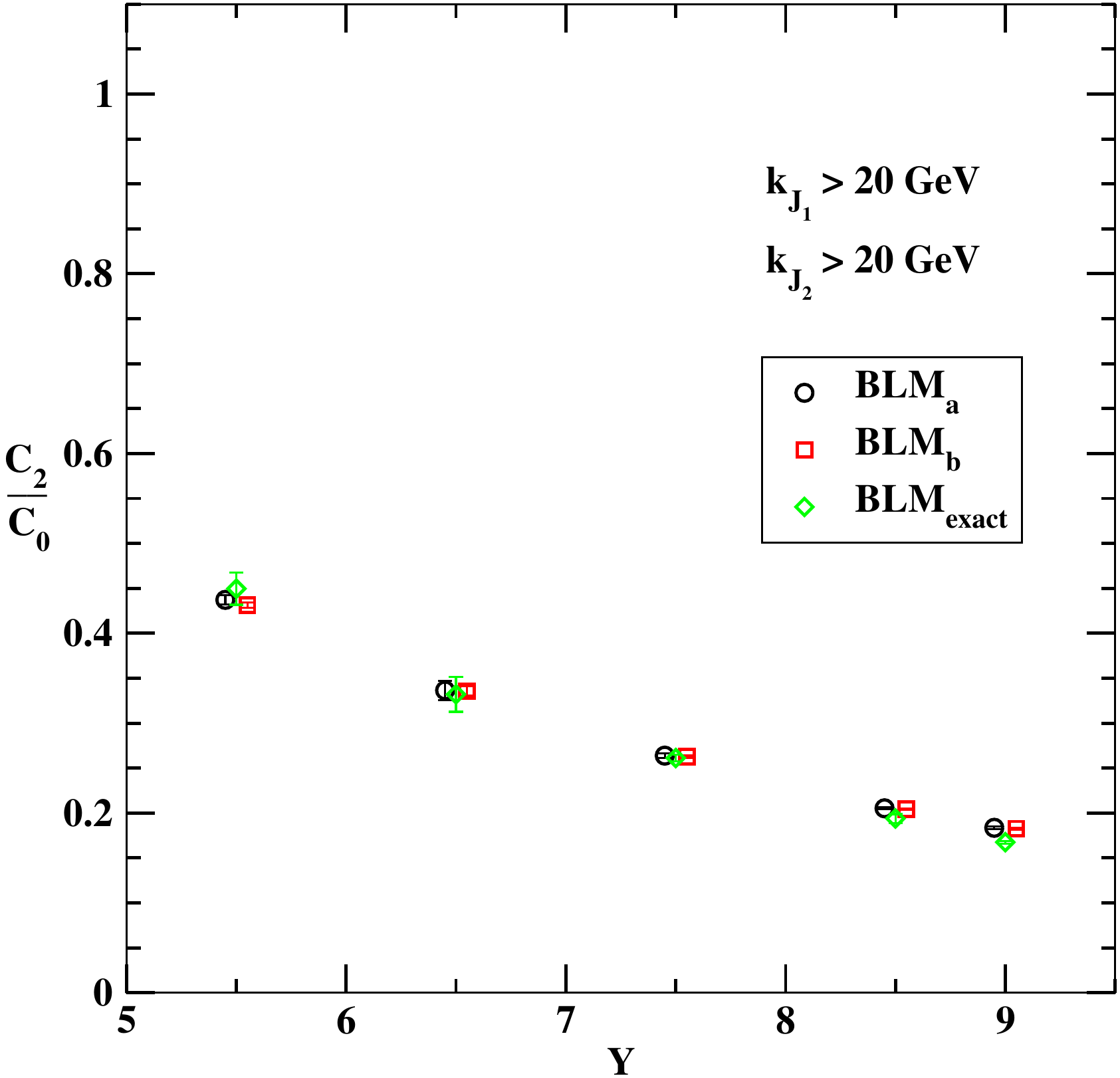}
   \includegraphics[scale=0.40]{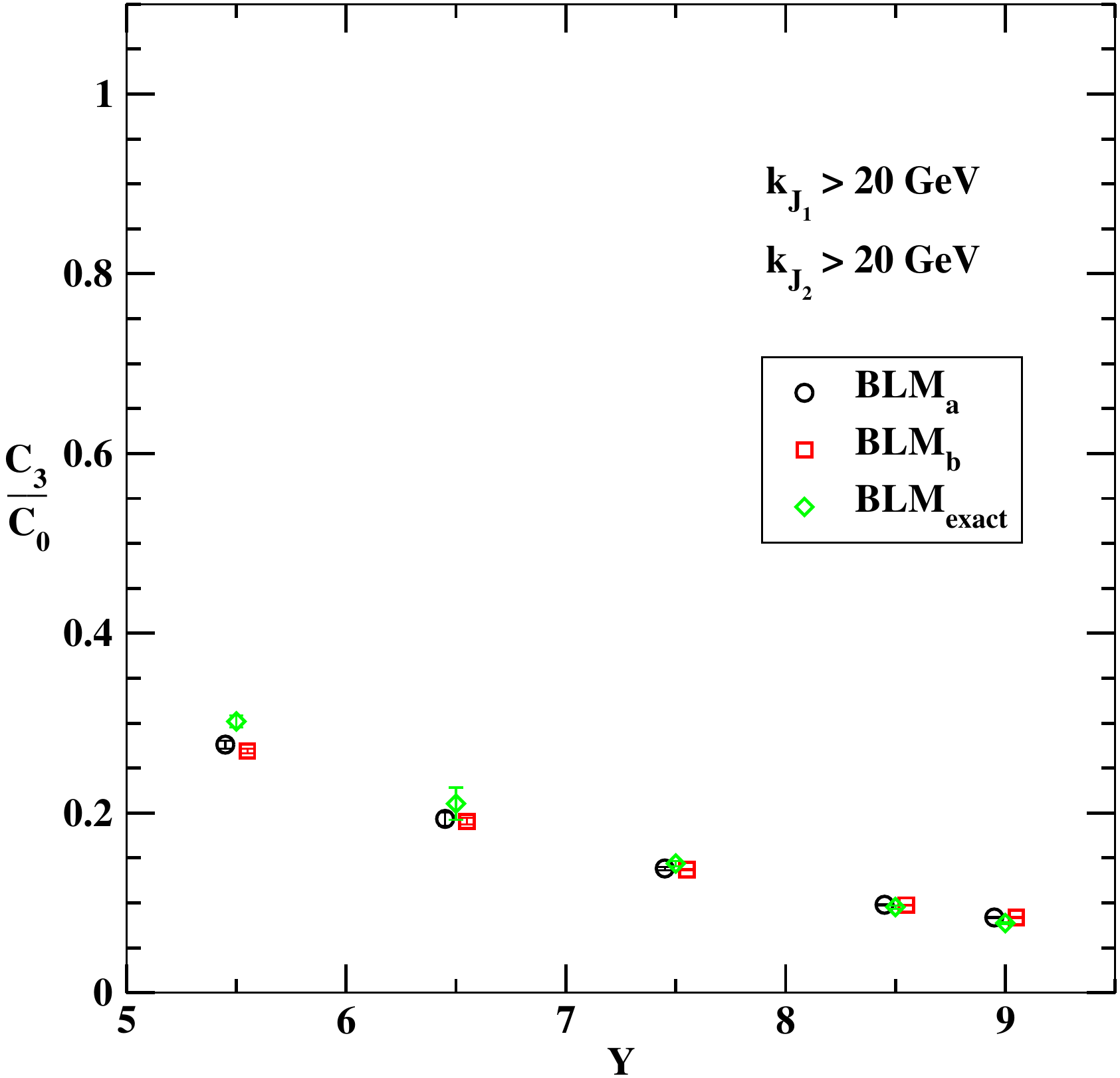}

   \includegraphics[scale=0.40]{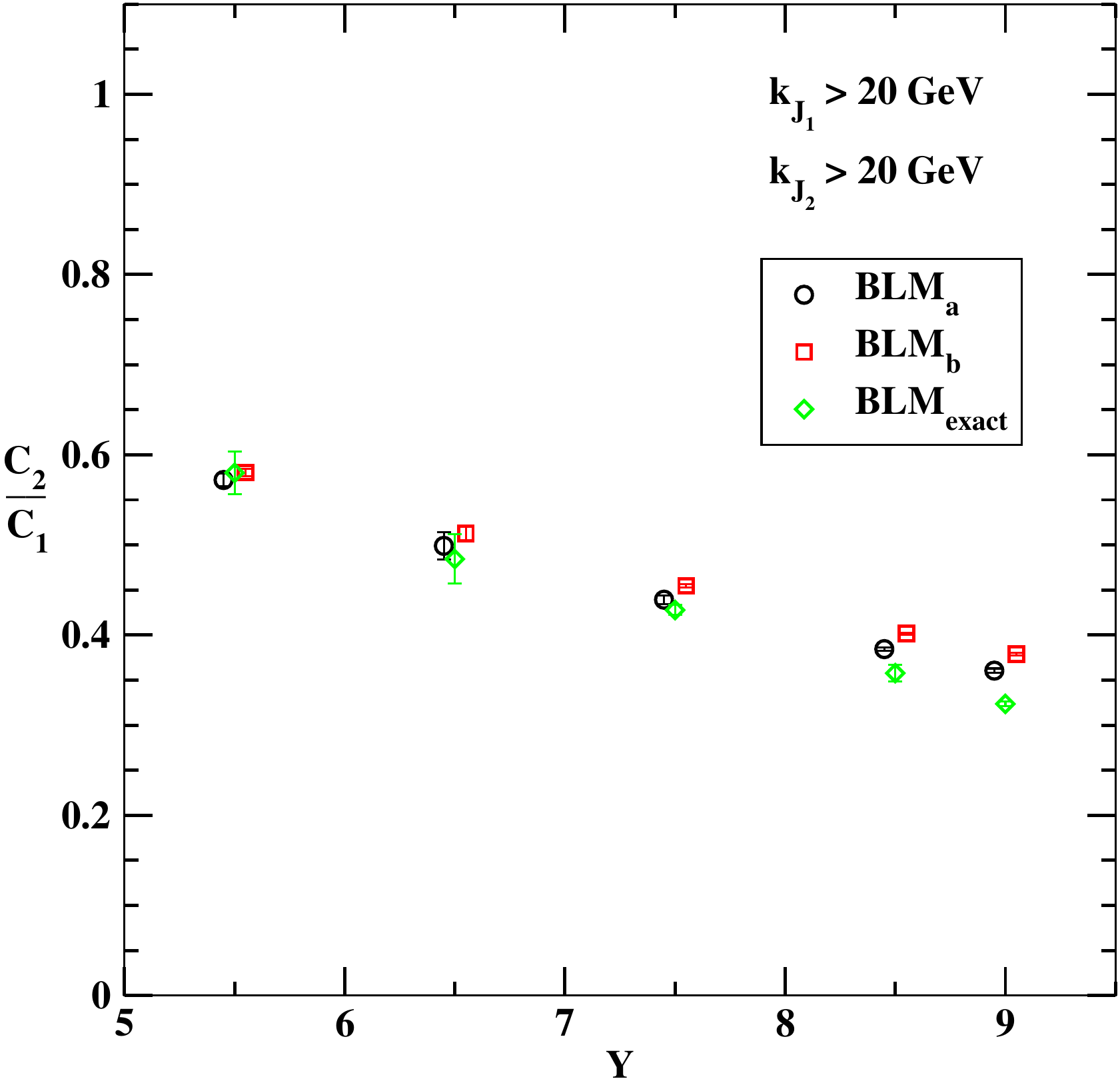}
   \includegraphics[scale=0.40]{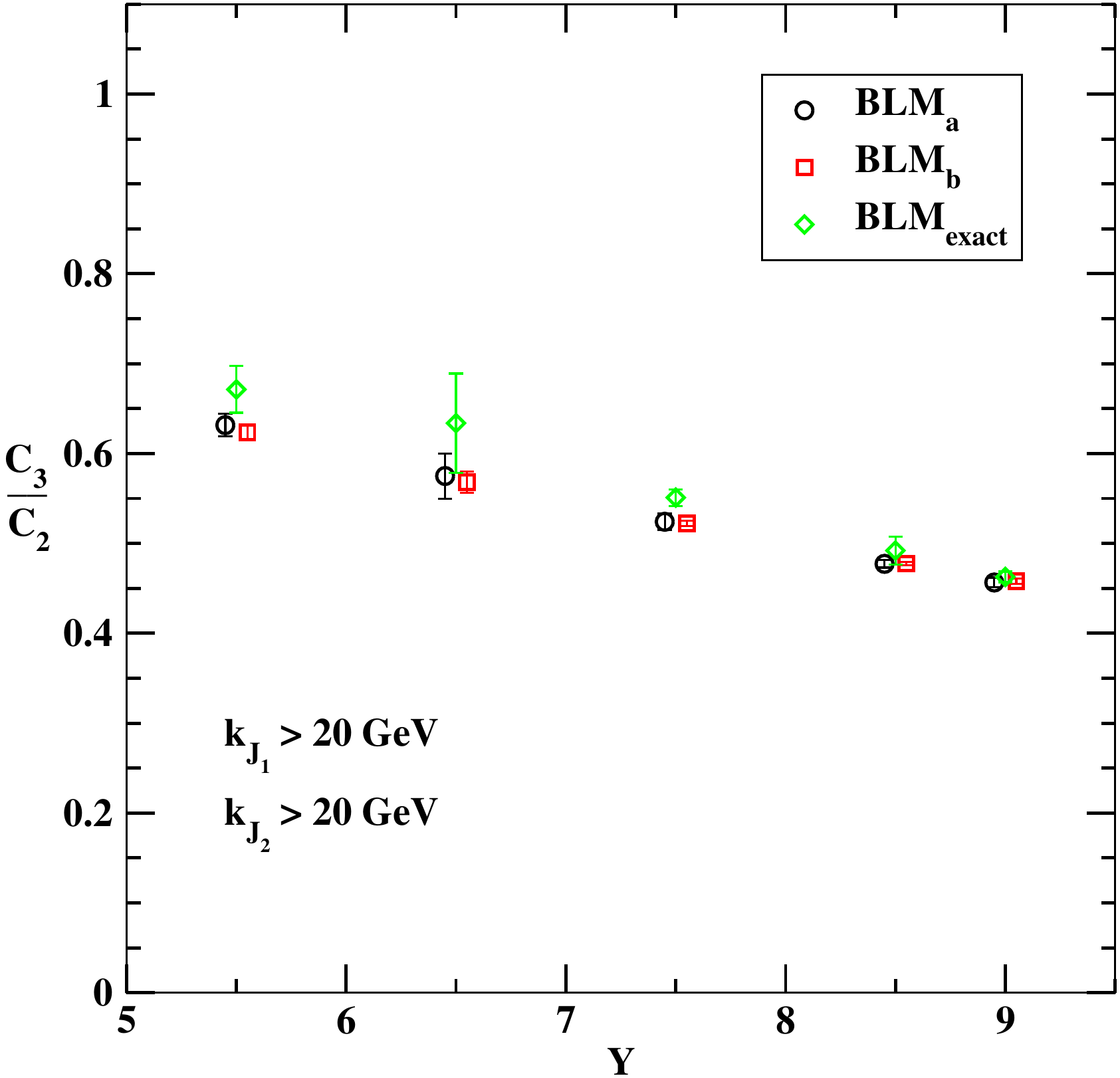}
\caption{$Y$-dependence of $C_0$ and of several ratios $C_m/C_n$
for $k_{J_1,\rm min}=k_{J_2,\rm min}=20$~GeV and for $|y_{J_1}| > 2.5$, from
the three variants of the BLM method (data points have been slightly shifted
along the horizontal axis for the sake of readability; see
Table~\ref{tab:2020_2.5}).}
\label{2020_2.5}
\end{figure}

\begin{figure}[p]
\centering

   \includegraphics[scale=0.40]{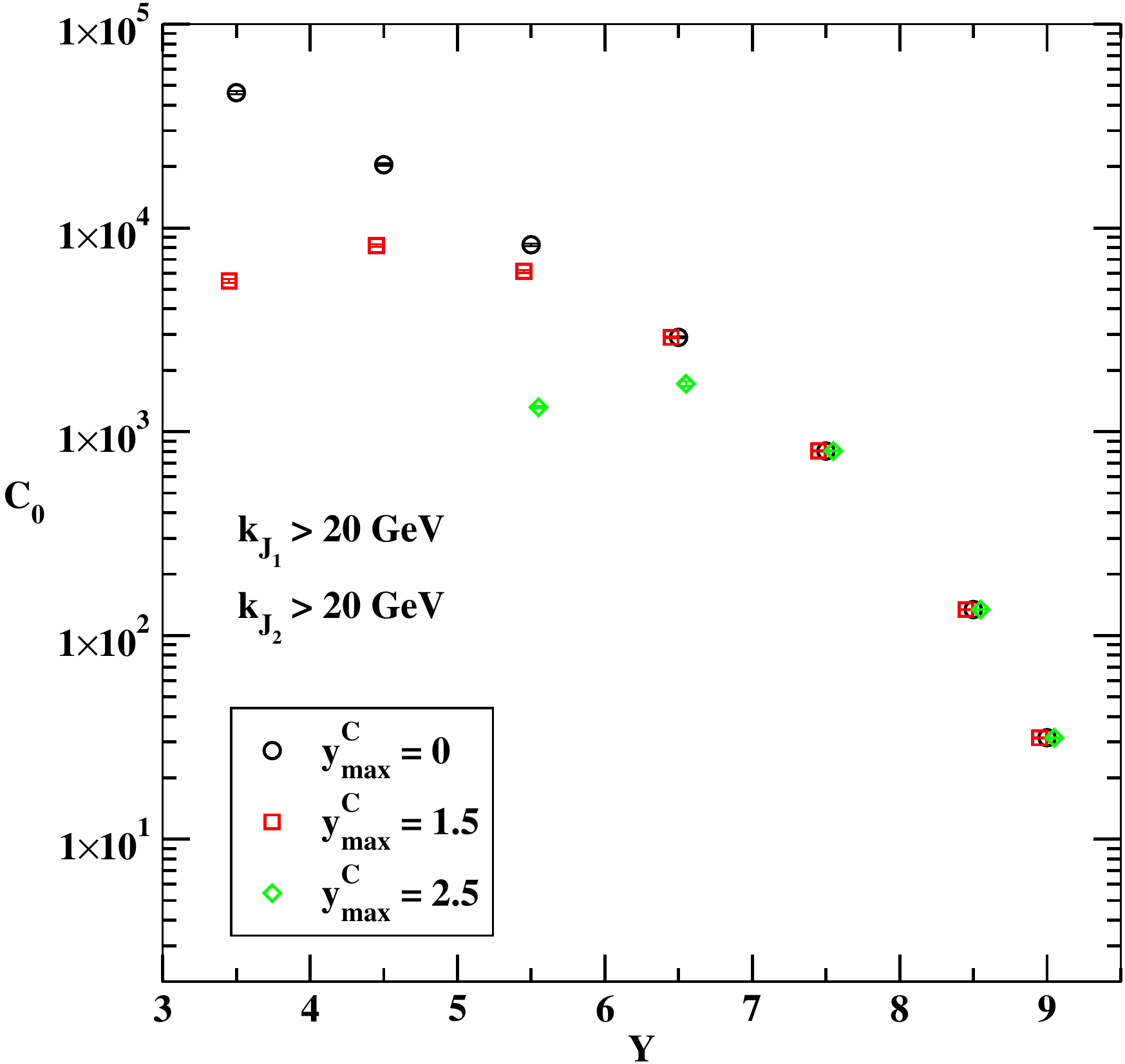}
   \includegraphics[scale=0.40]{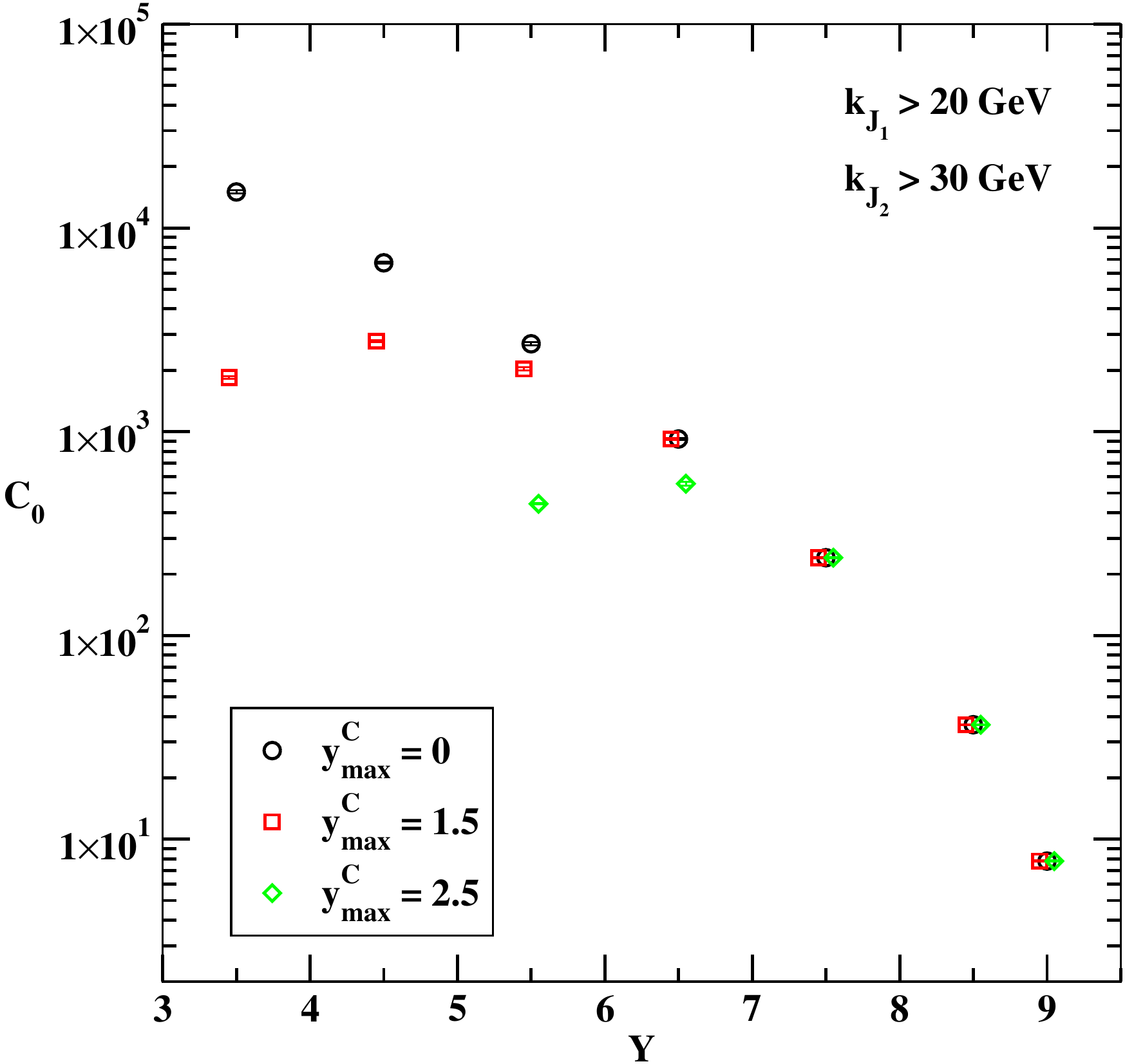}

   \includegraphics[scale=0.40]{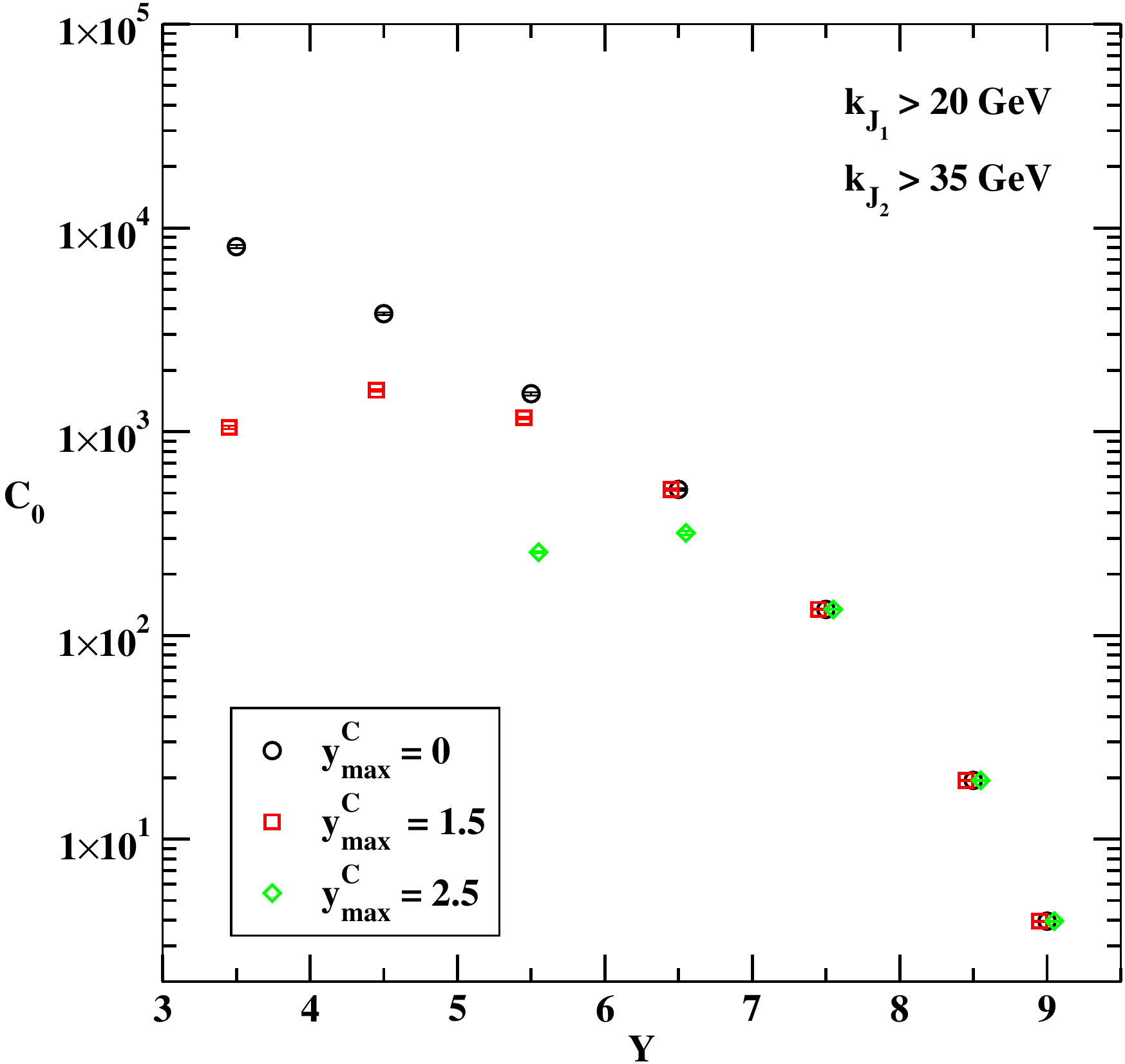}
   \includegraphics[scale=0.40]{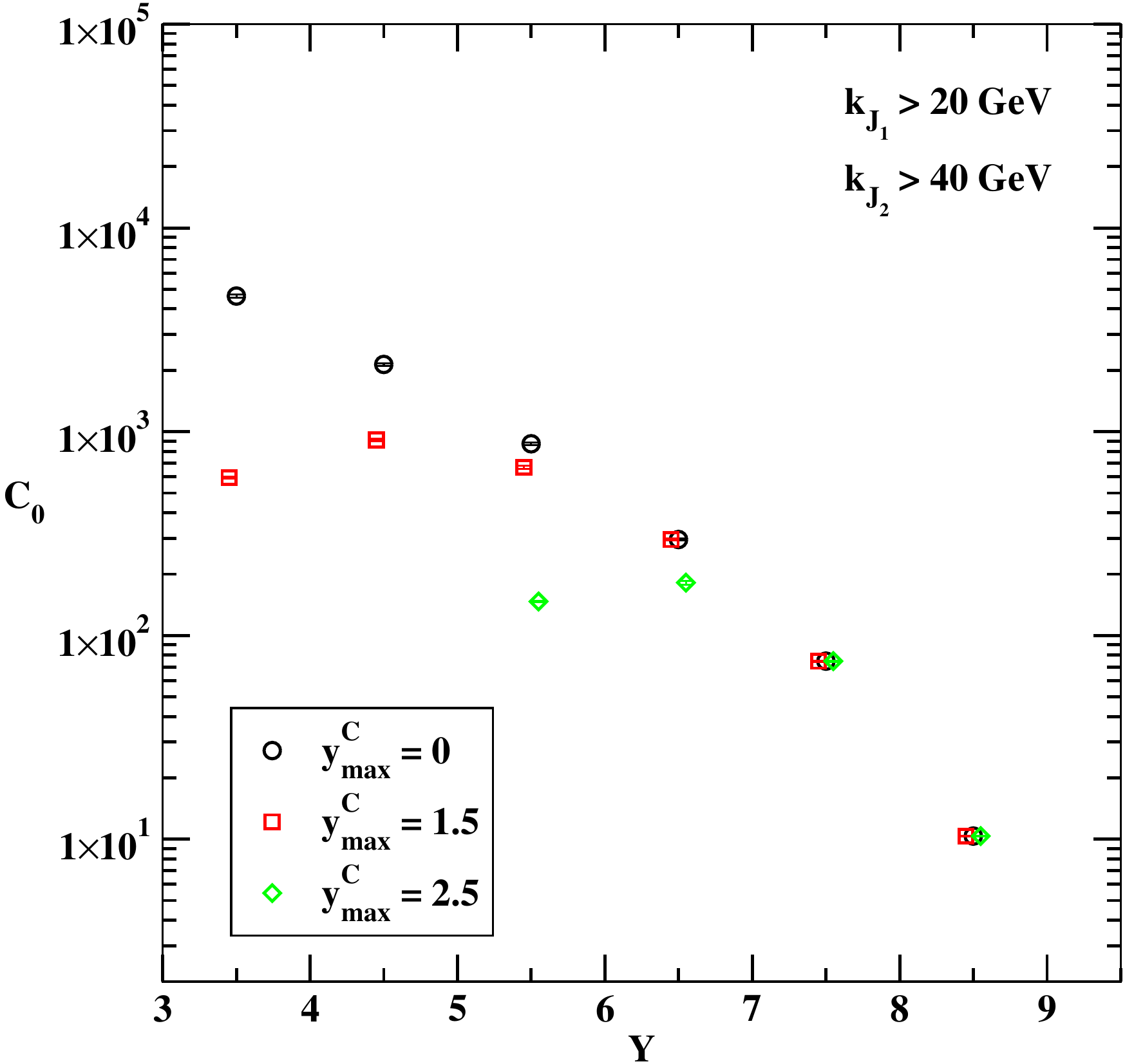}

   \includegraphics[scale=0.40]{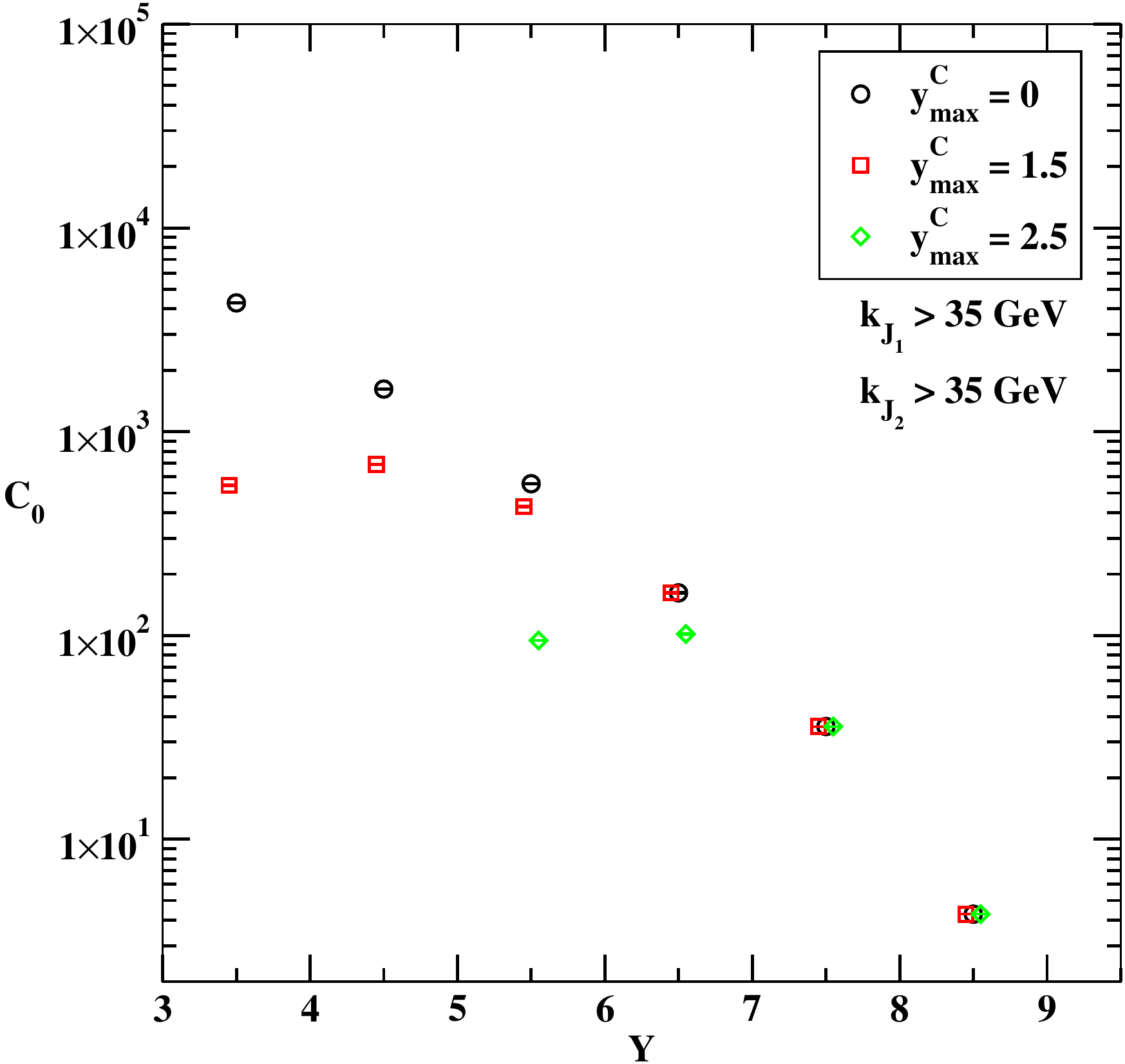}
\caption{$Y$-dependence of $C_0$ from the ``exact'' BLM method, for all
choices of the cuts on jet transverse momenta and of the central rapidity region
(data points have been slightly shifted along the horizontal axis
for the sake of readability; see Table~\ref{tab:C0_e}).}
\label{C0_e}
\end{figure}

\begin{figure}[p]
\centering

   \includegraphics[scale=0.40]{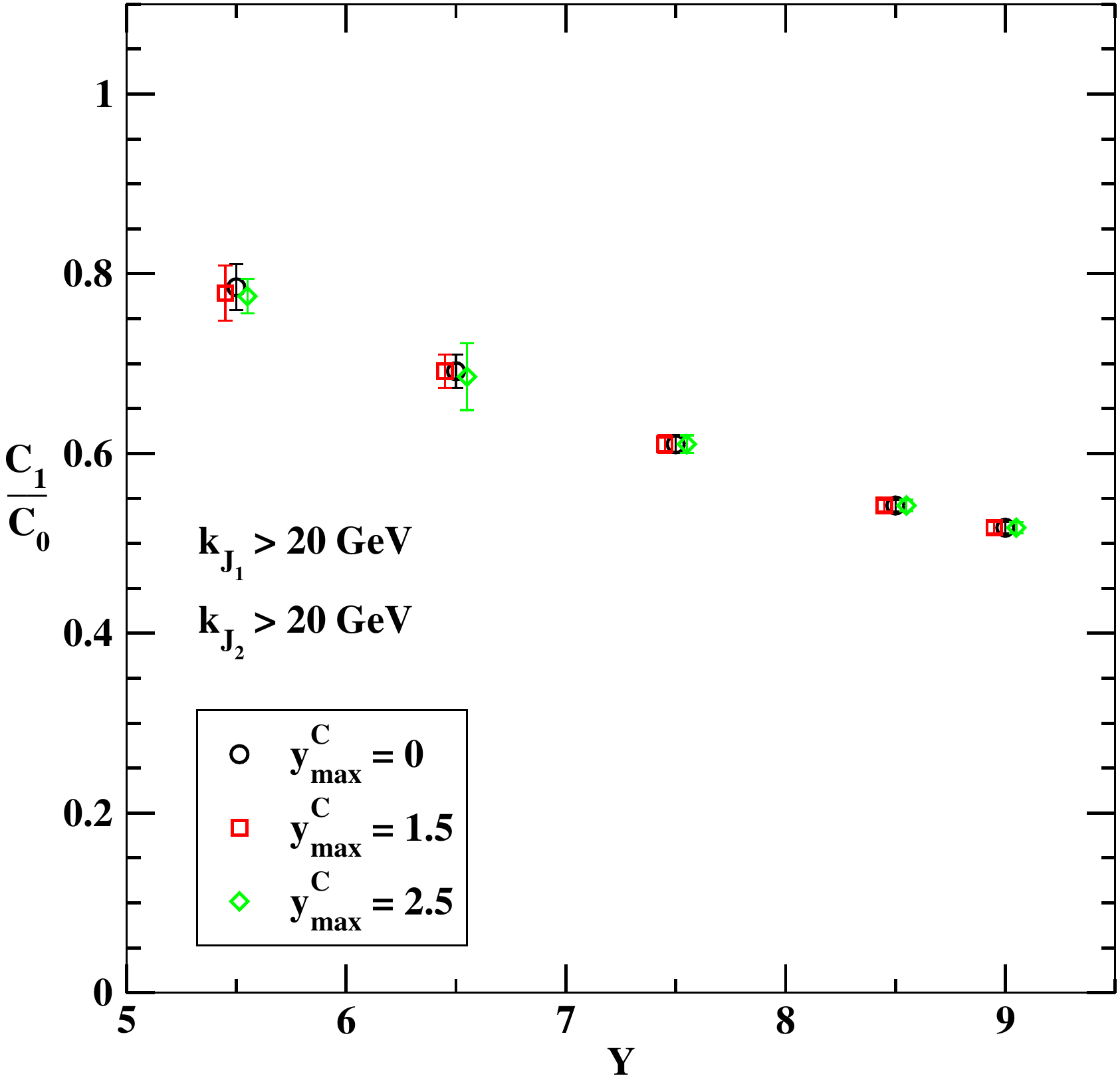}
   \includegraphics[scale=0.40]{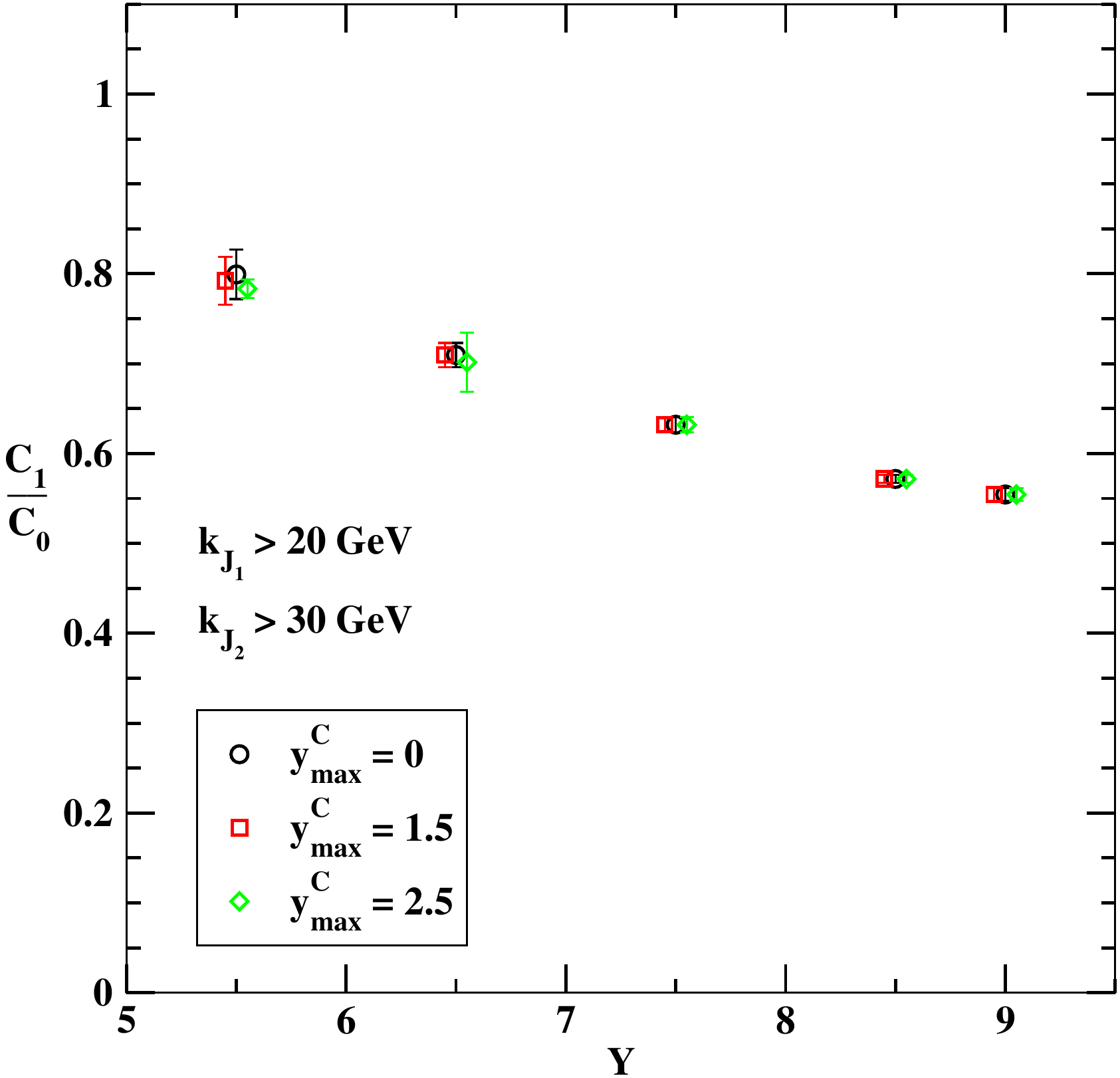}

   \includegraphics[scale=0.40]{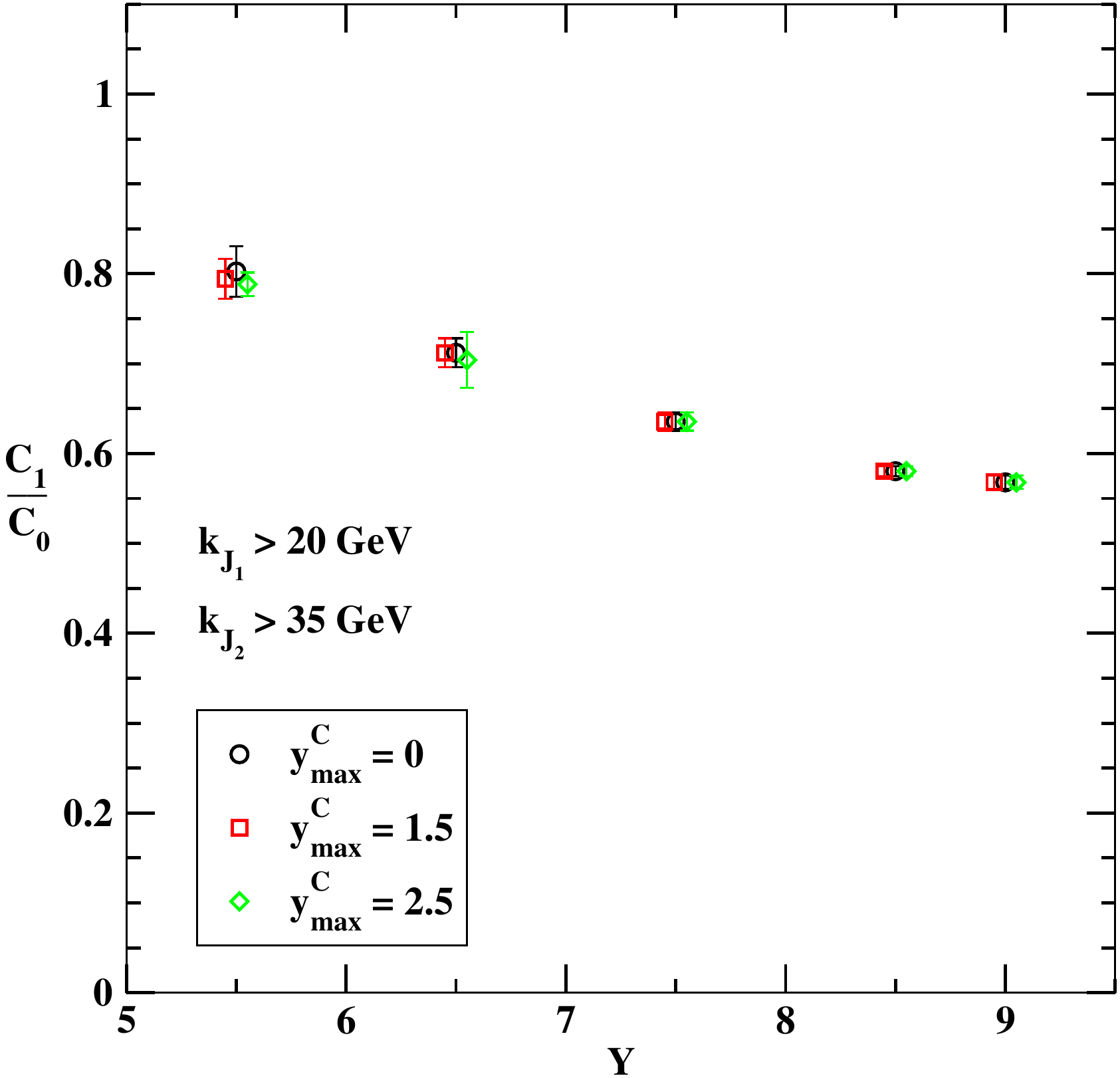}
   \includegraphics[scale=0.40]{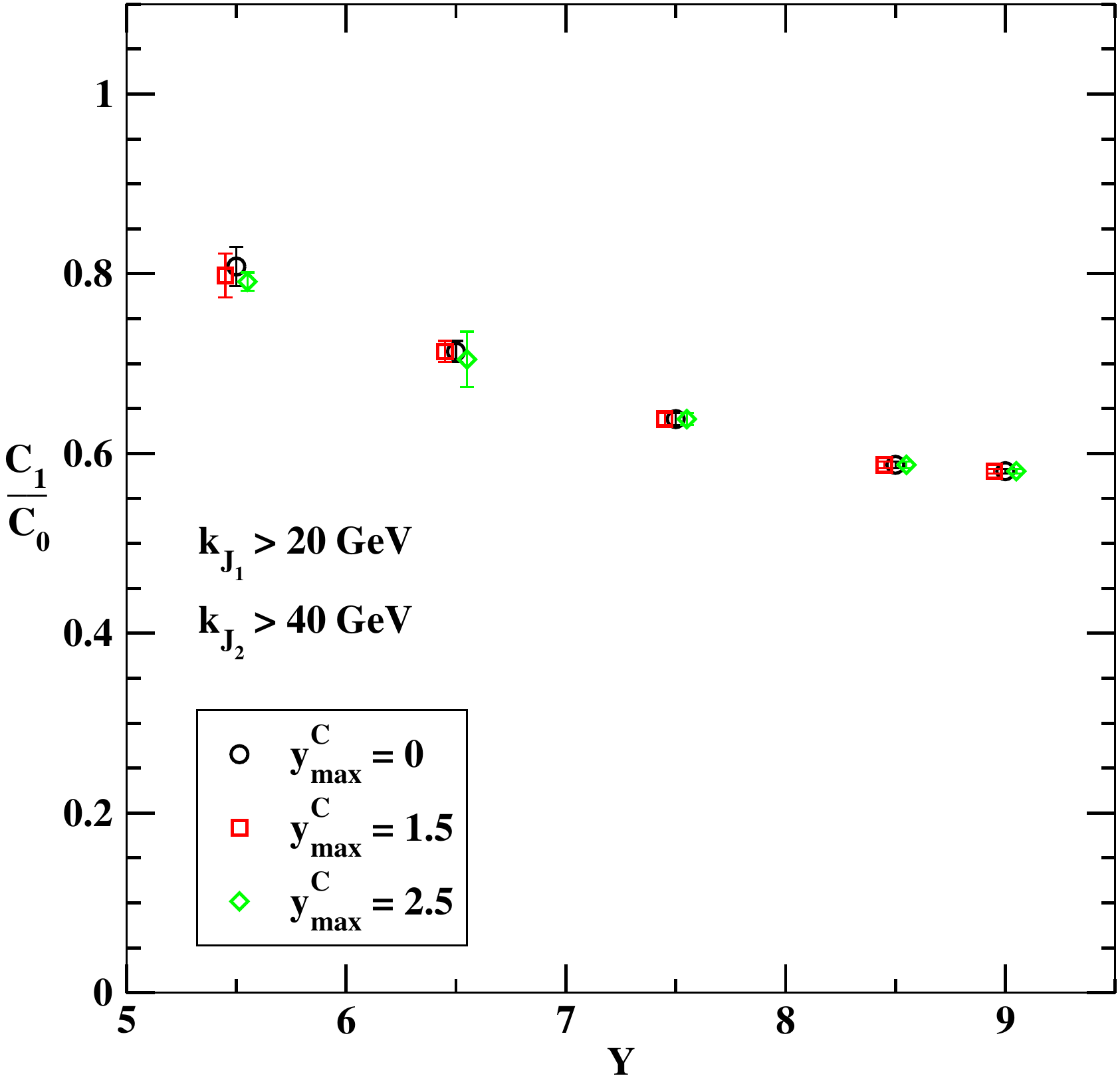}

   \includegraphics[scale=0.40]{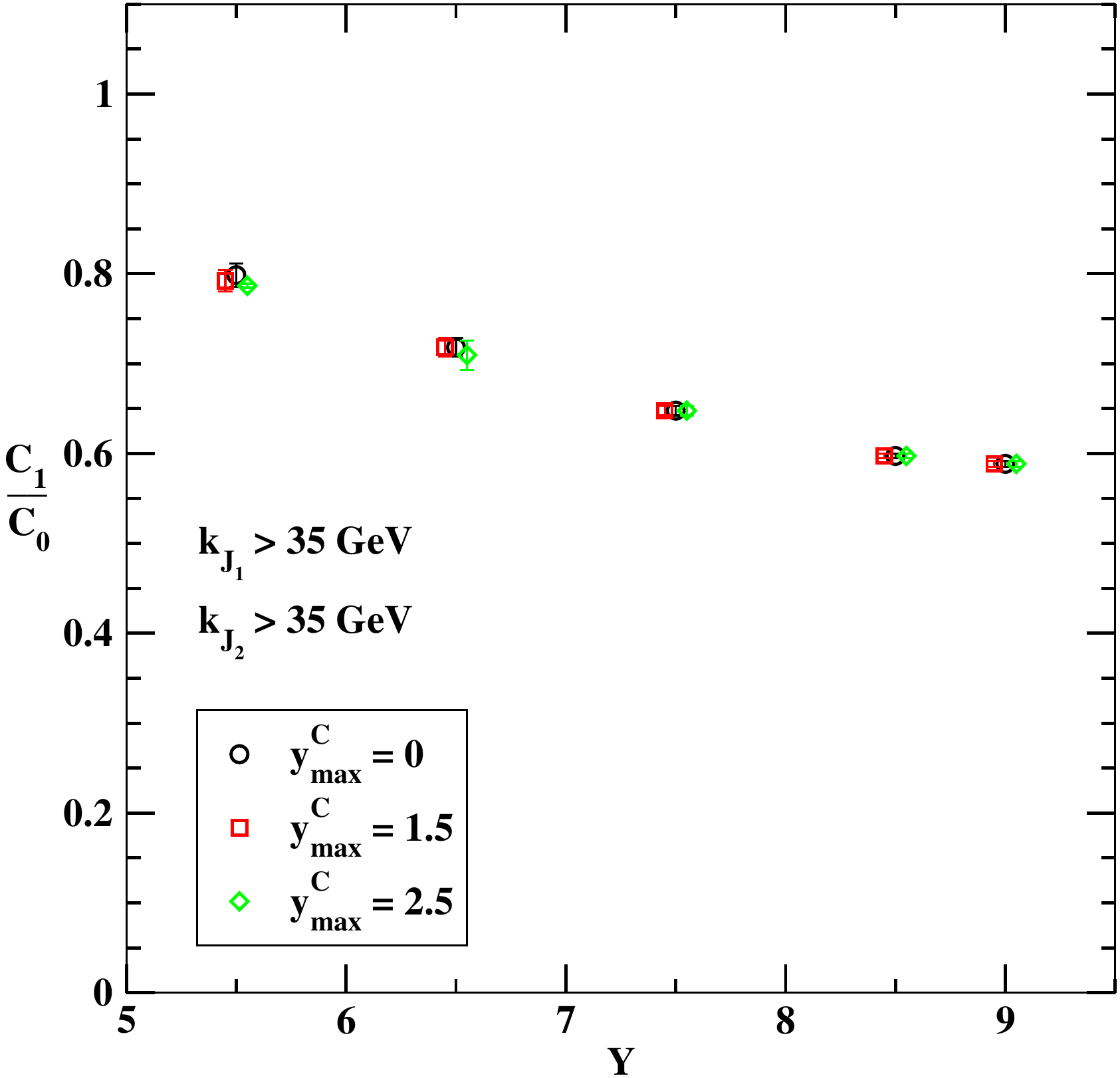}
\caption{$Y$-dependence of $C_1/C_0$ from the ``exact'' BLM method,
for all choices of the cuts on jet transverse momenta and of the central
rapidity region (data points have been slightly shifted along the horizontal
axis for the sake of readability; see Table~\ref{tab:C1C0_e}).}
\label{C1C0_e}
\end{figure}

\begin{figure}[p]
\centering

   \includegraphics[scale=0.40]{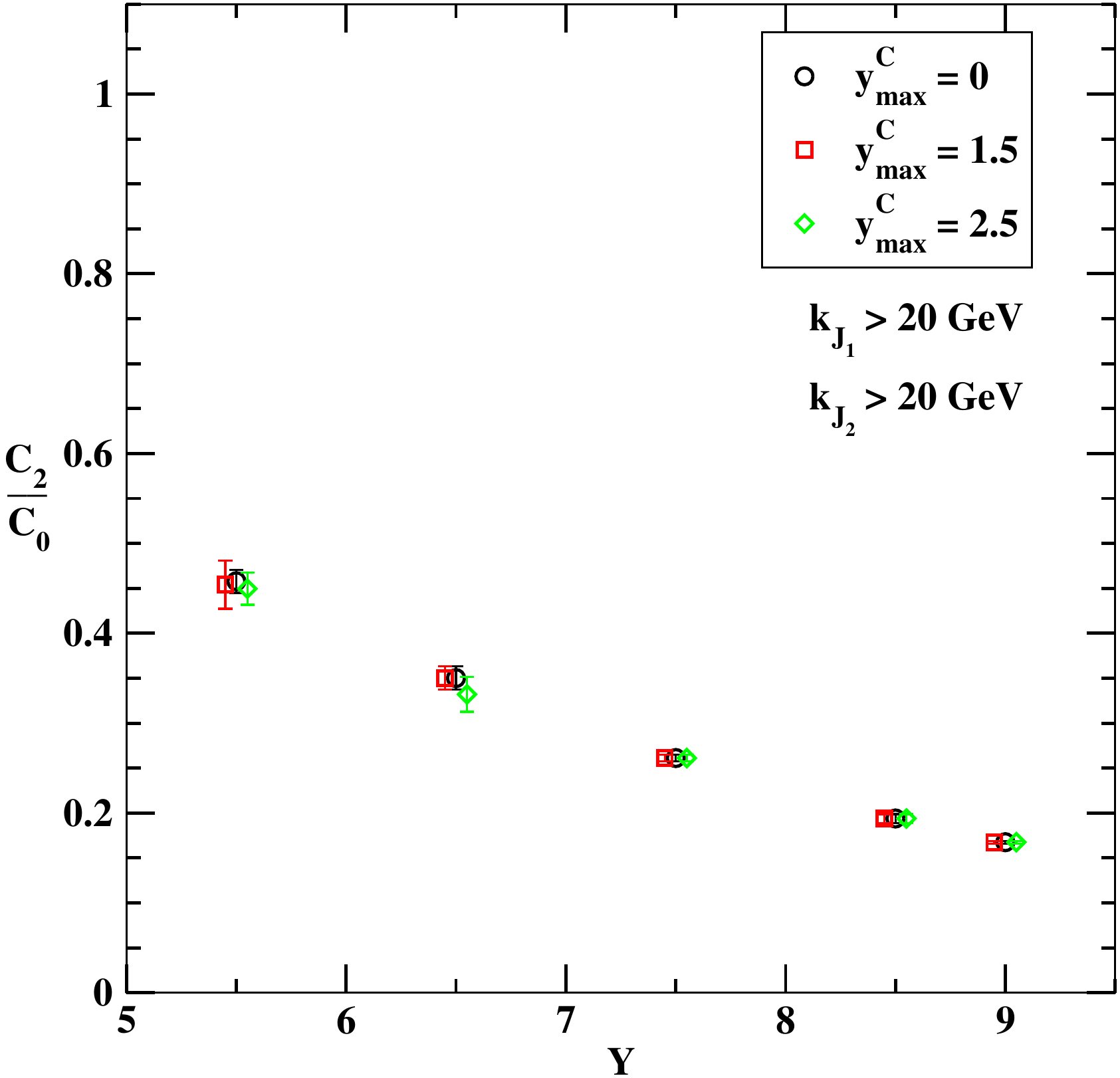}
   \includegraphics[scale=0.40]{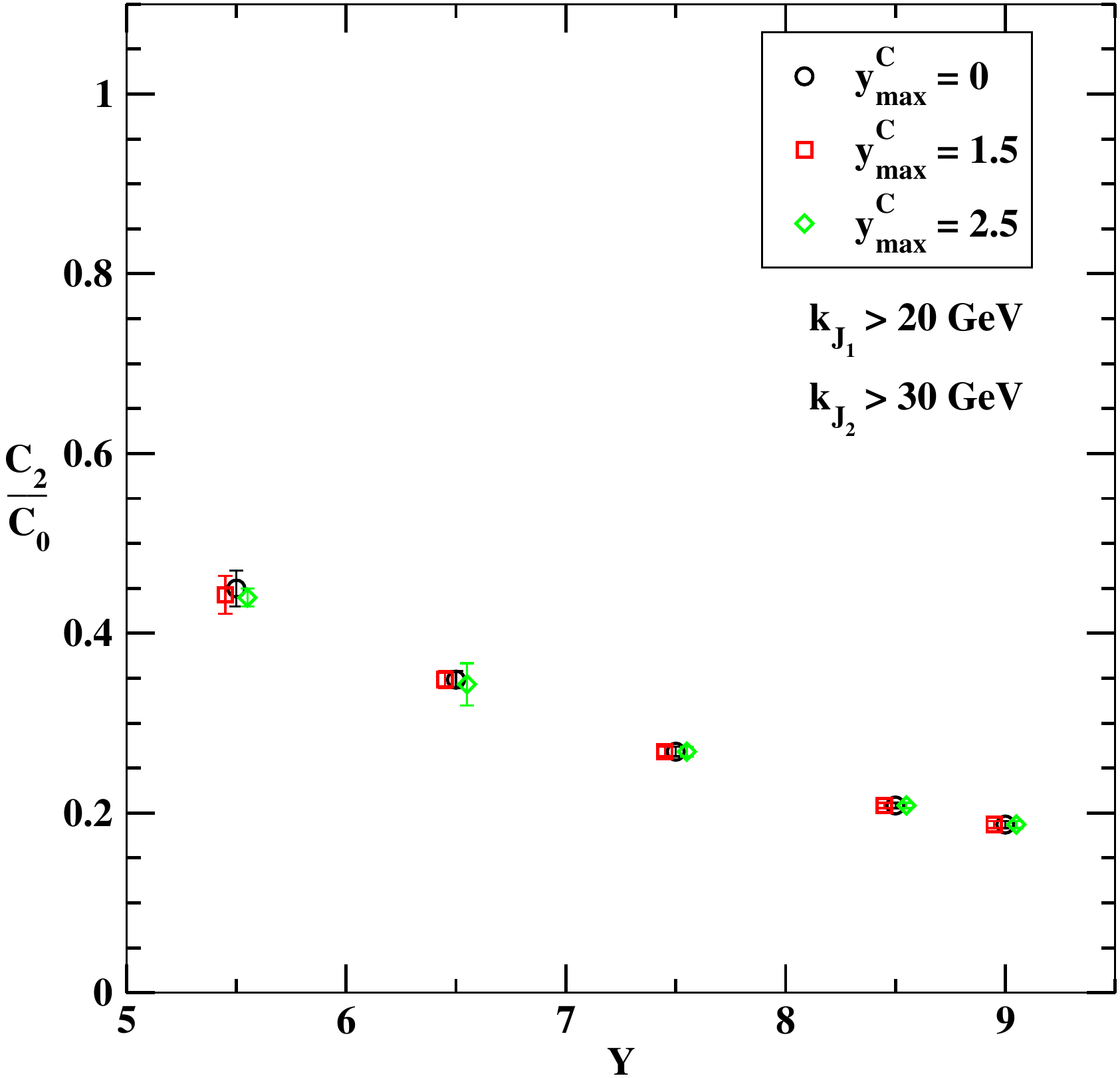}

   \includegraphics[scale=0.40]{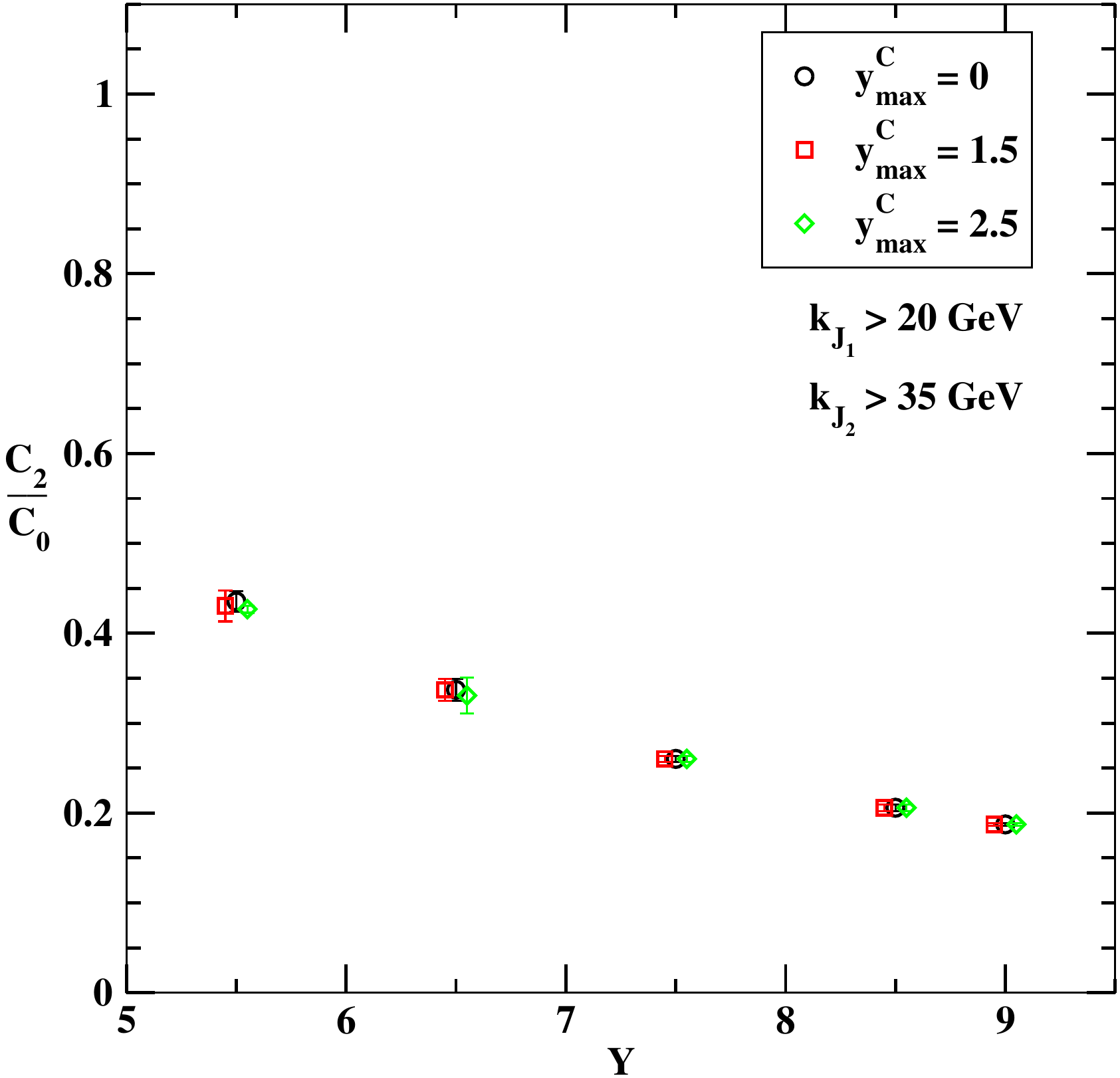}
   \includegraphics[scale=0.40]{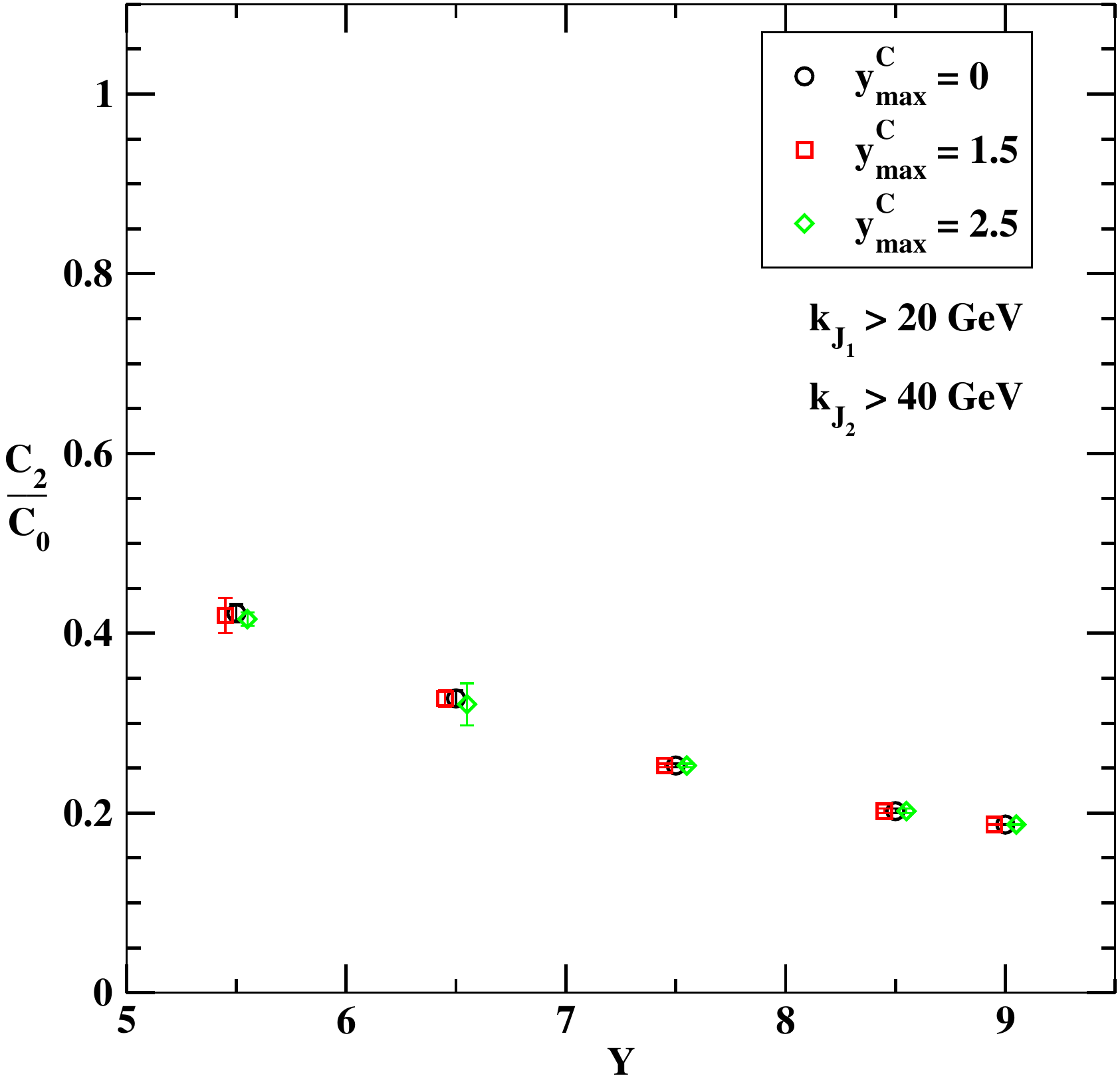}

   \includegraphics[scale=0.40]{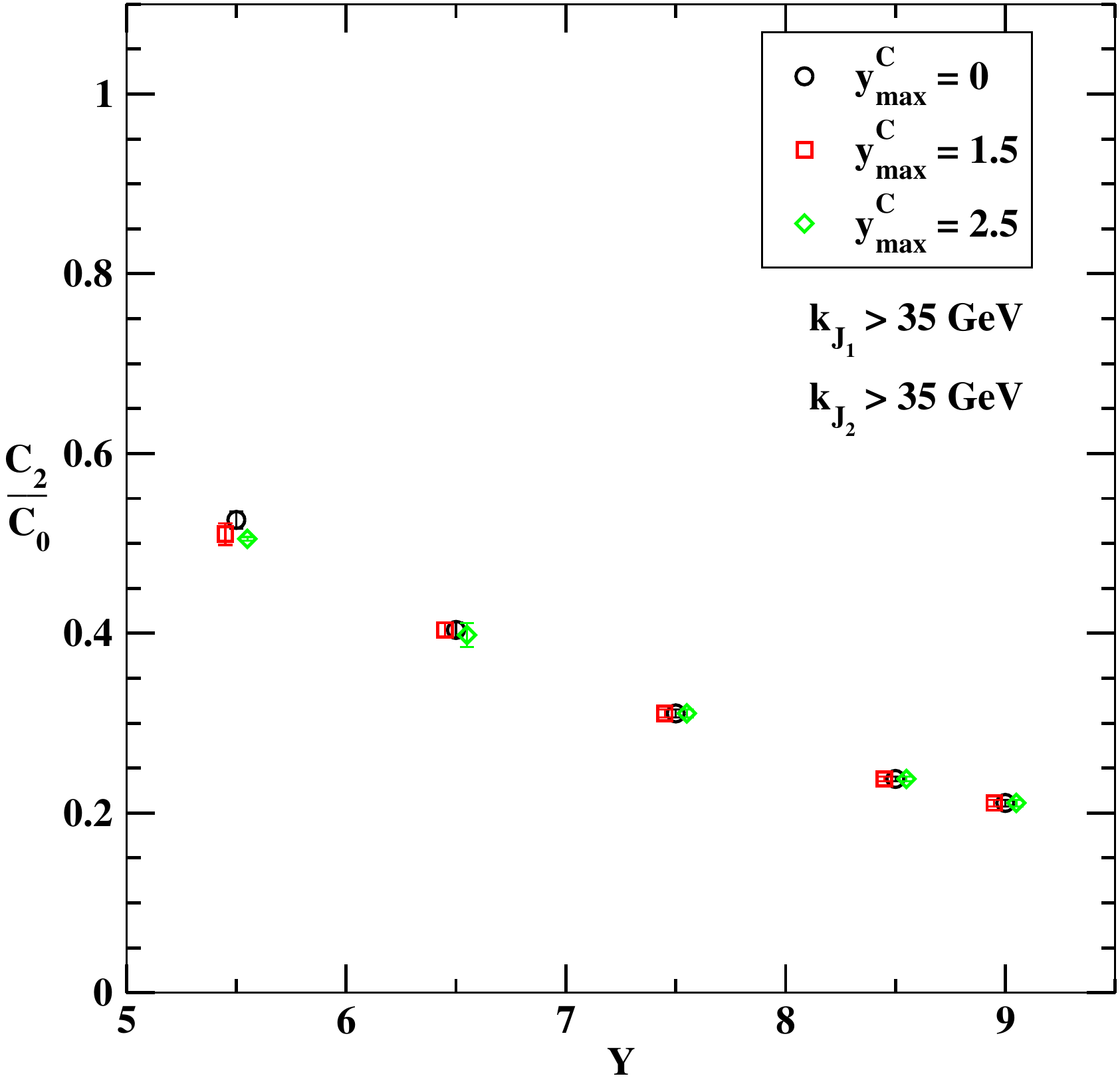}
\caption{$Y$-dependence of $C_2/C_0$ from the ``exact'' BLM method,
for all choices of the cuts on jet transverse momenta and of the central
rapidity region (data points have been slightly shifted along the horizontal
axis for the sake of readability; see Table~\ref{tab:C2C0_e}).}
\label{C2C0_e}
\end{figure}

\begin{figure}[p]
\centering

   \includegraphics[scale=0.40]{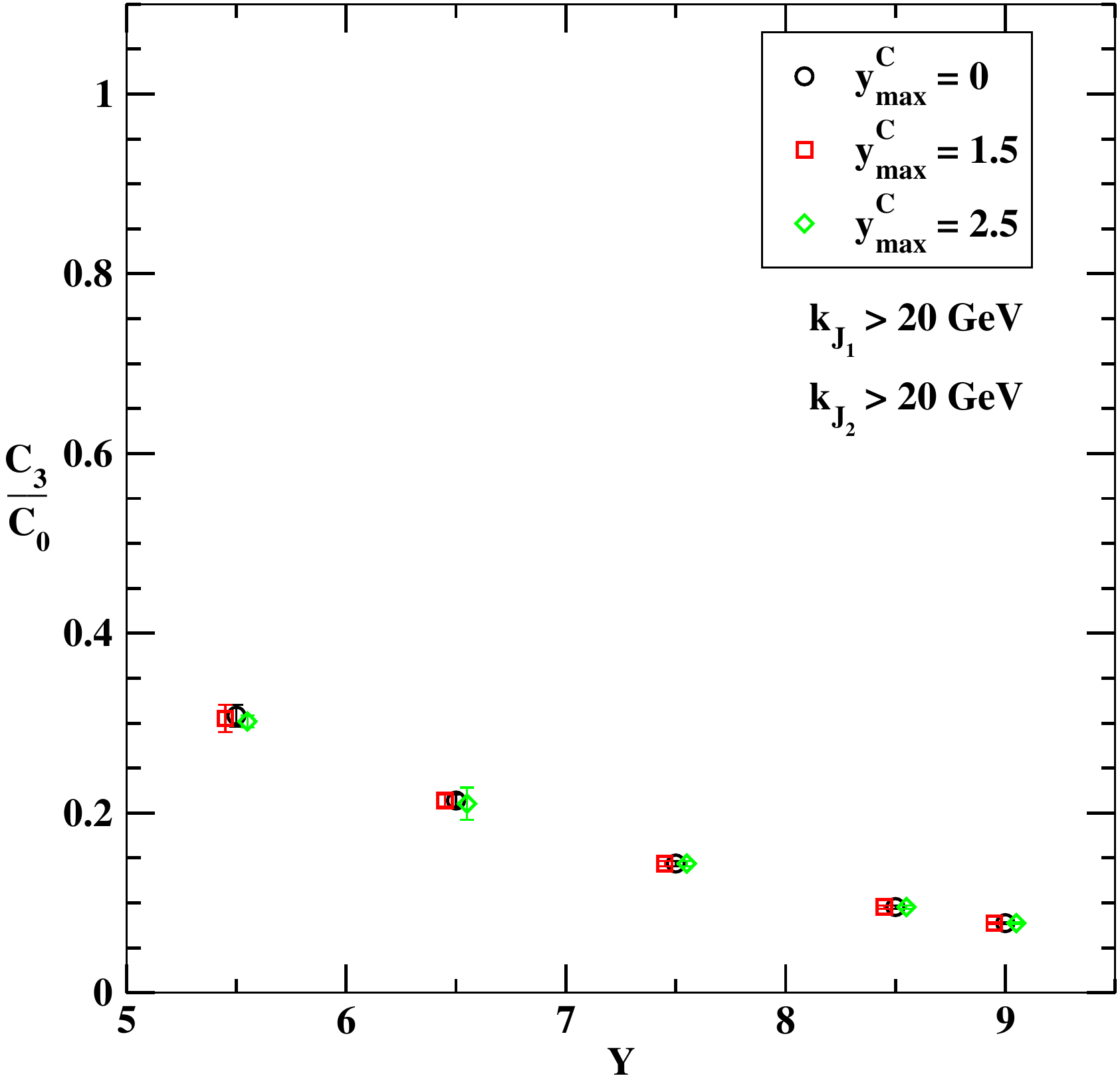}
   \includegraphics[scale=0.40]{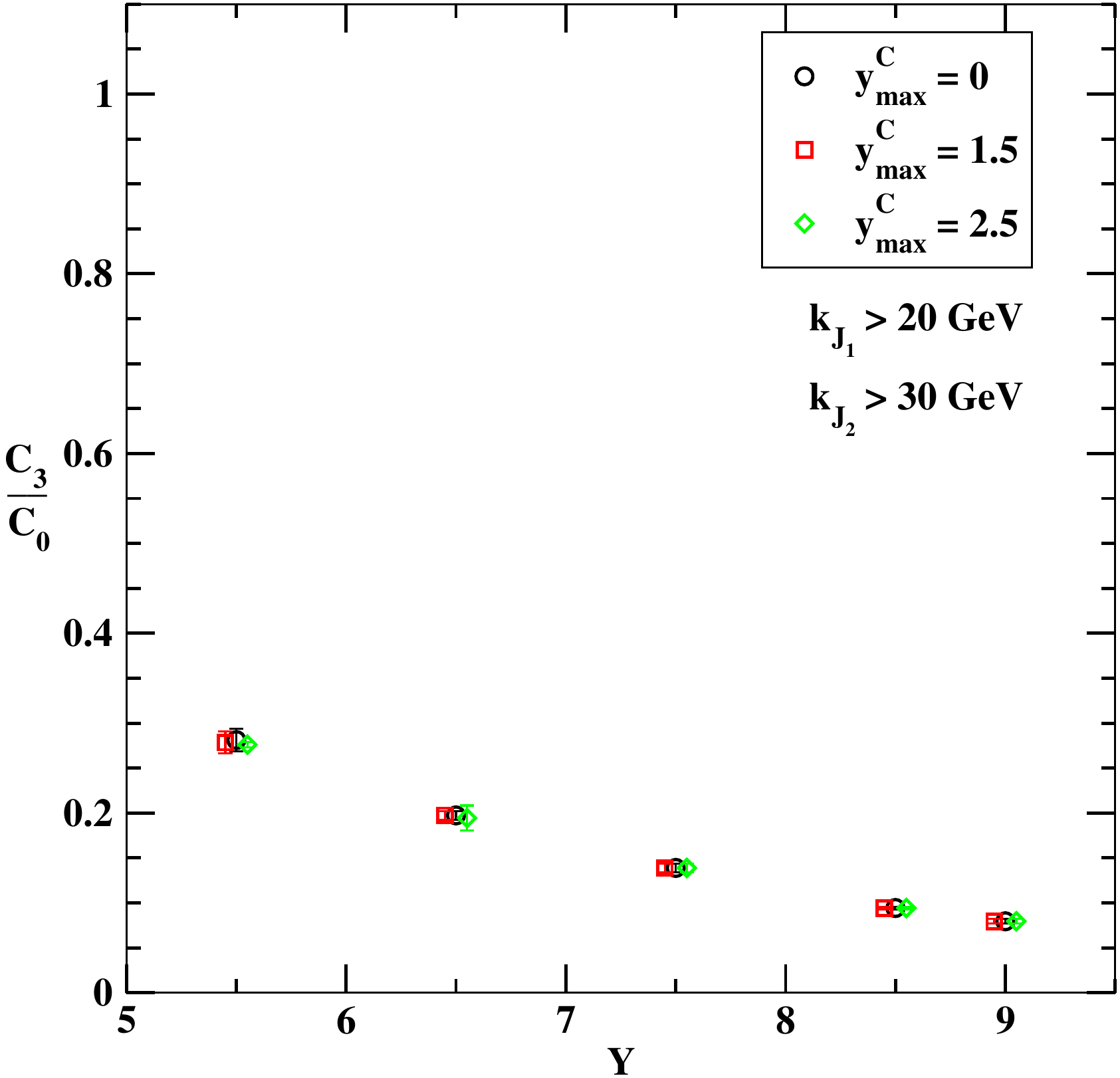}

   \includegraphics[scale=0.40]{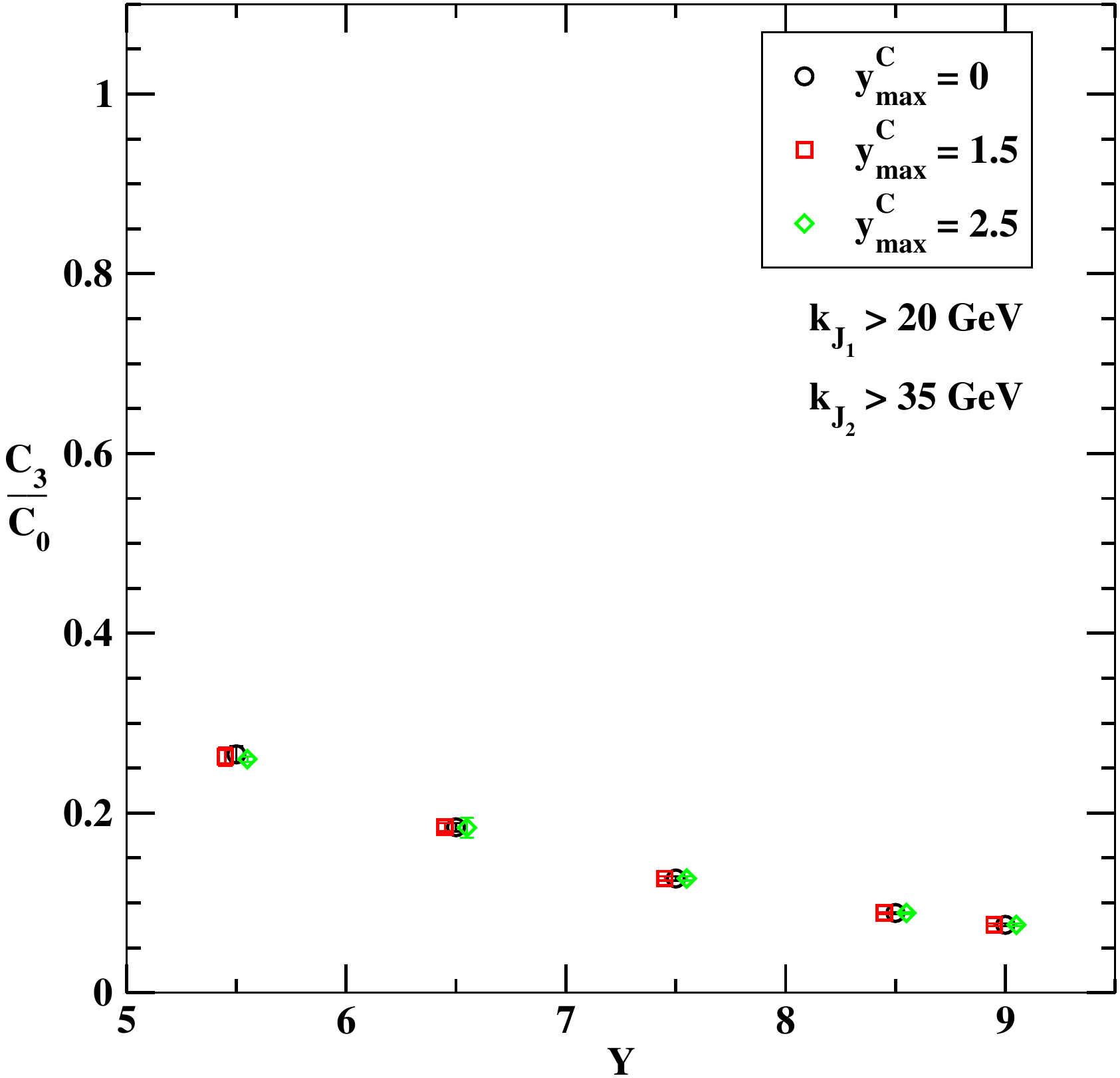}
   \includegraphics[scale=0.40]{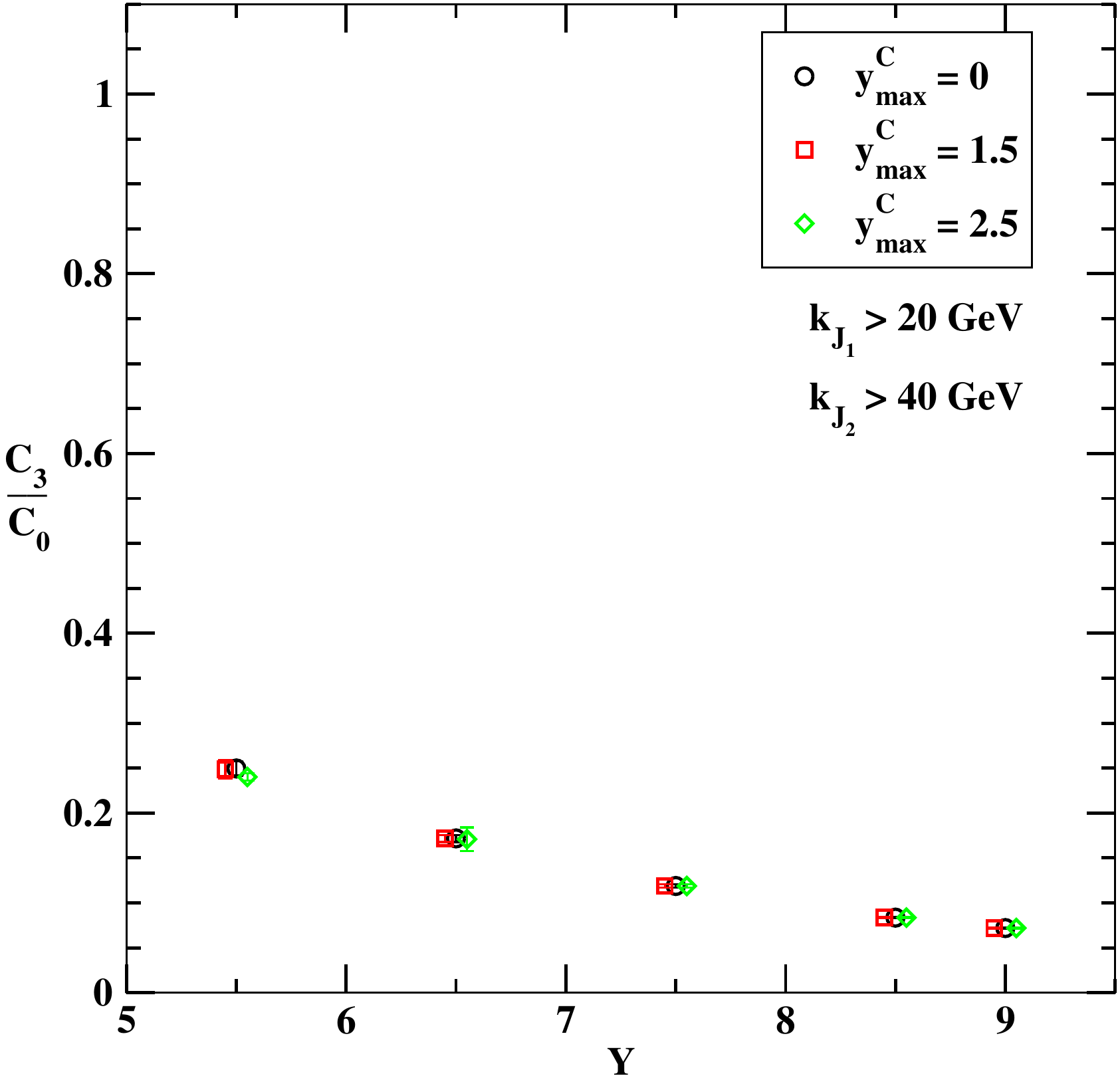}

   \includegraphics[scale=0.40]{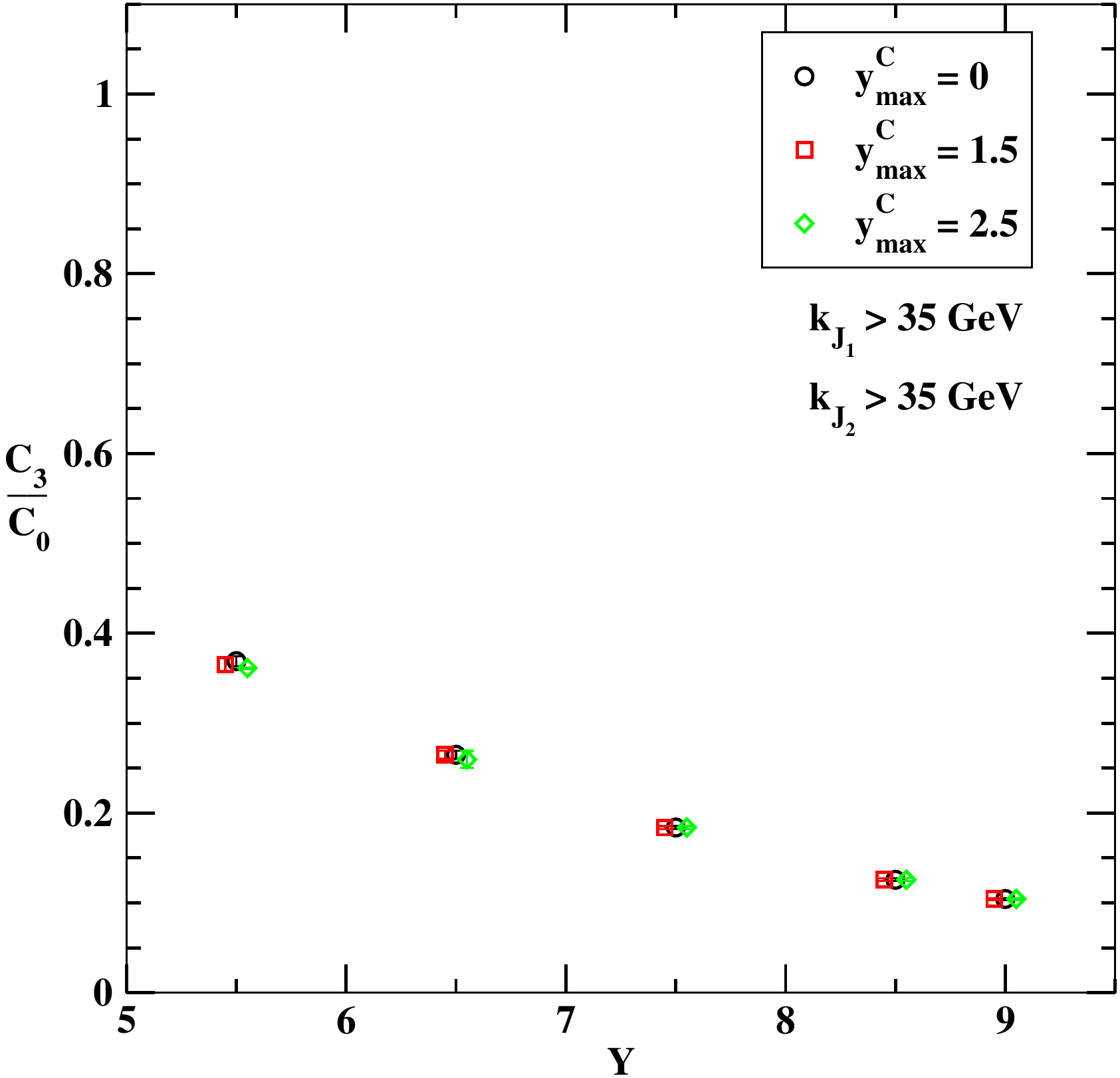}
\caption{$Y$-dependence of $C_3/C_0$ from the ``exact'' BLM method,
for all choices of the cuts on jet transverse momenta and of the central
rapidity region (data points have been slightly shifted along the horizontal
axis for the sake of readability; see Table~\ref{tab:C3C0_e}).}
\label{C3C0_e}
\end{figure}

\begin{figure}[p]
\centering

   \includegraphics[scale=0.40]{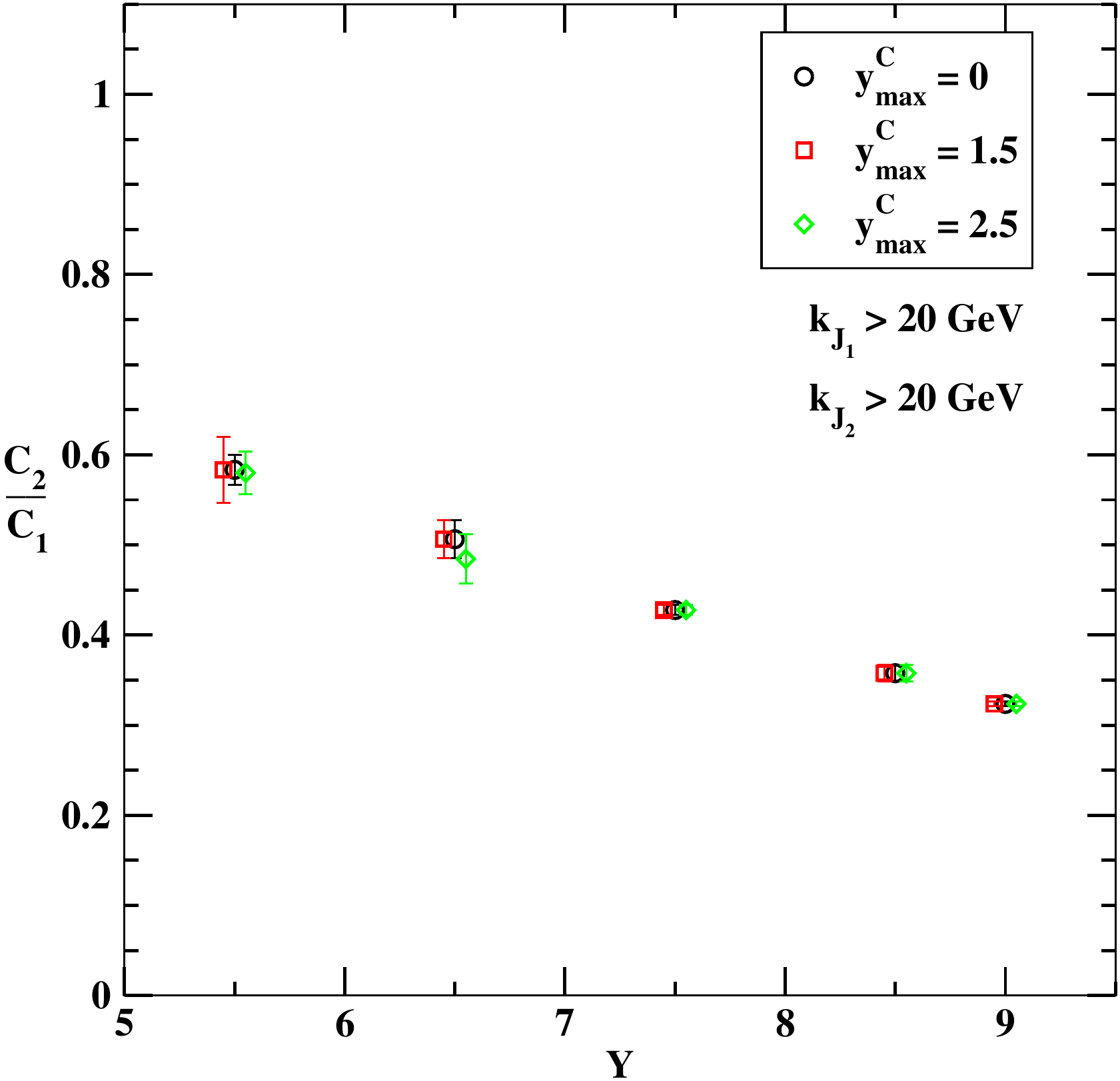}
   \includegraphics[scale=0.40]{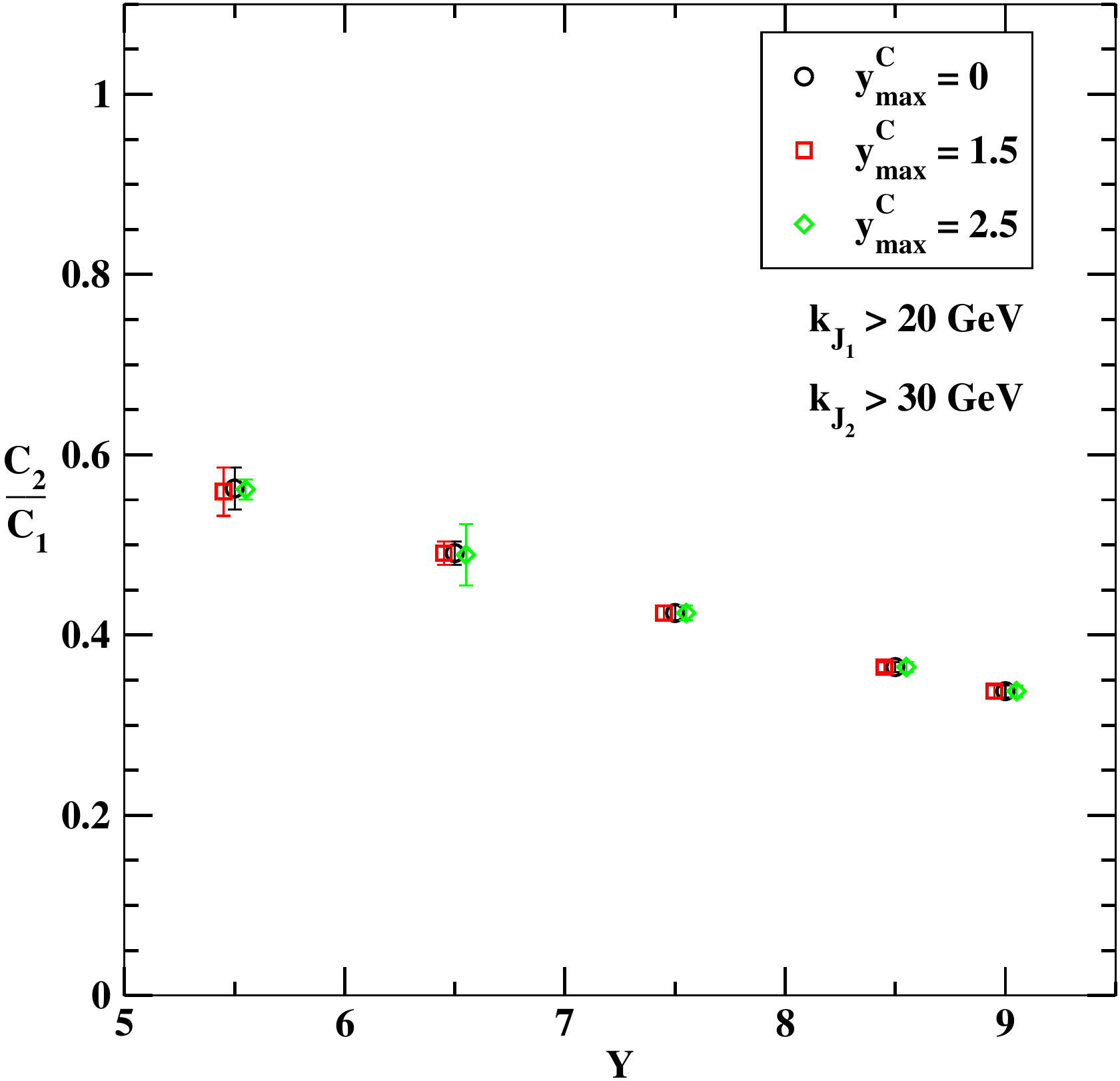}

   \includegraphics[scale=0.40]{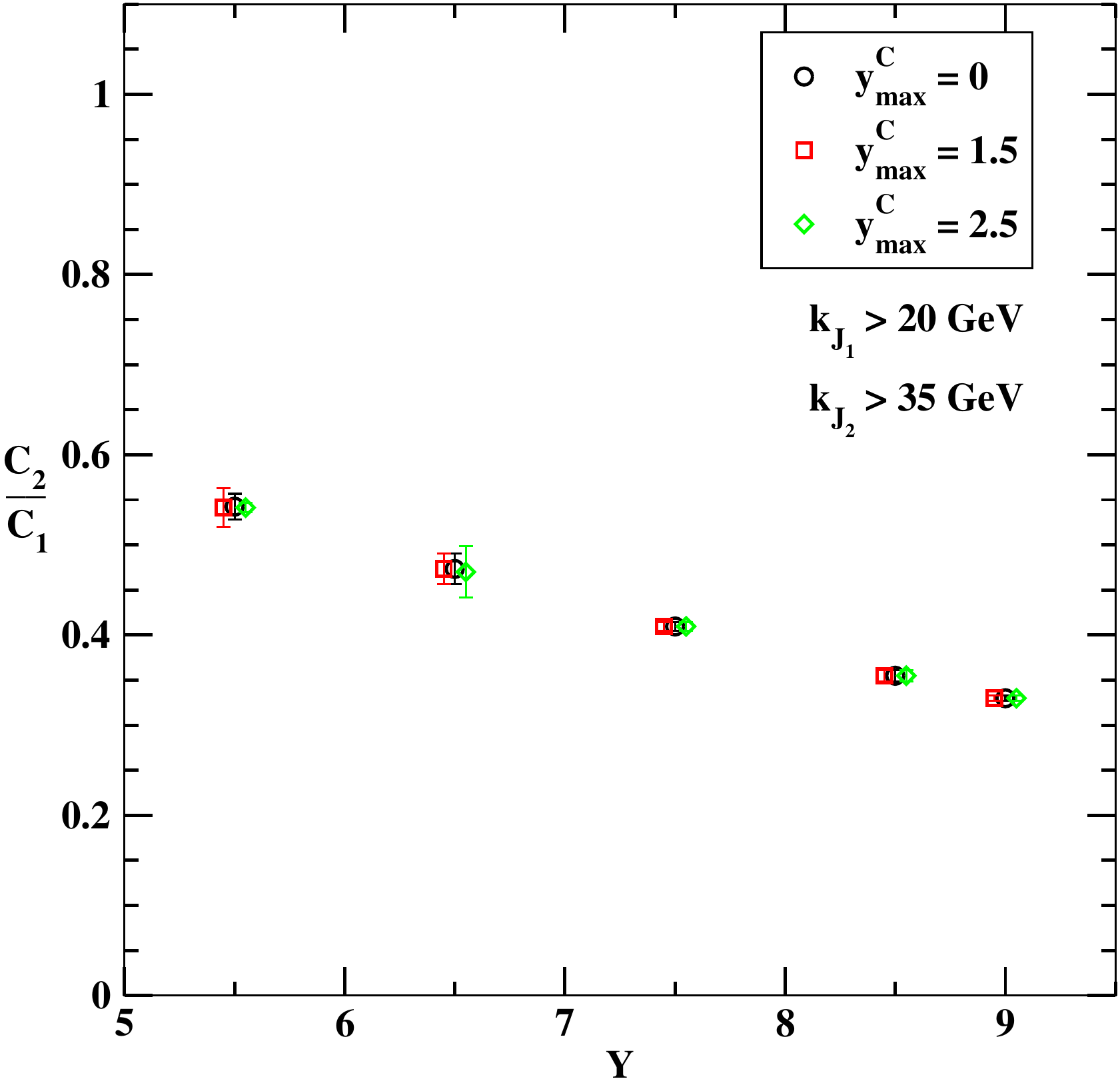}
   \includegraphics[scale=0.40]{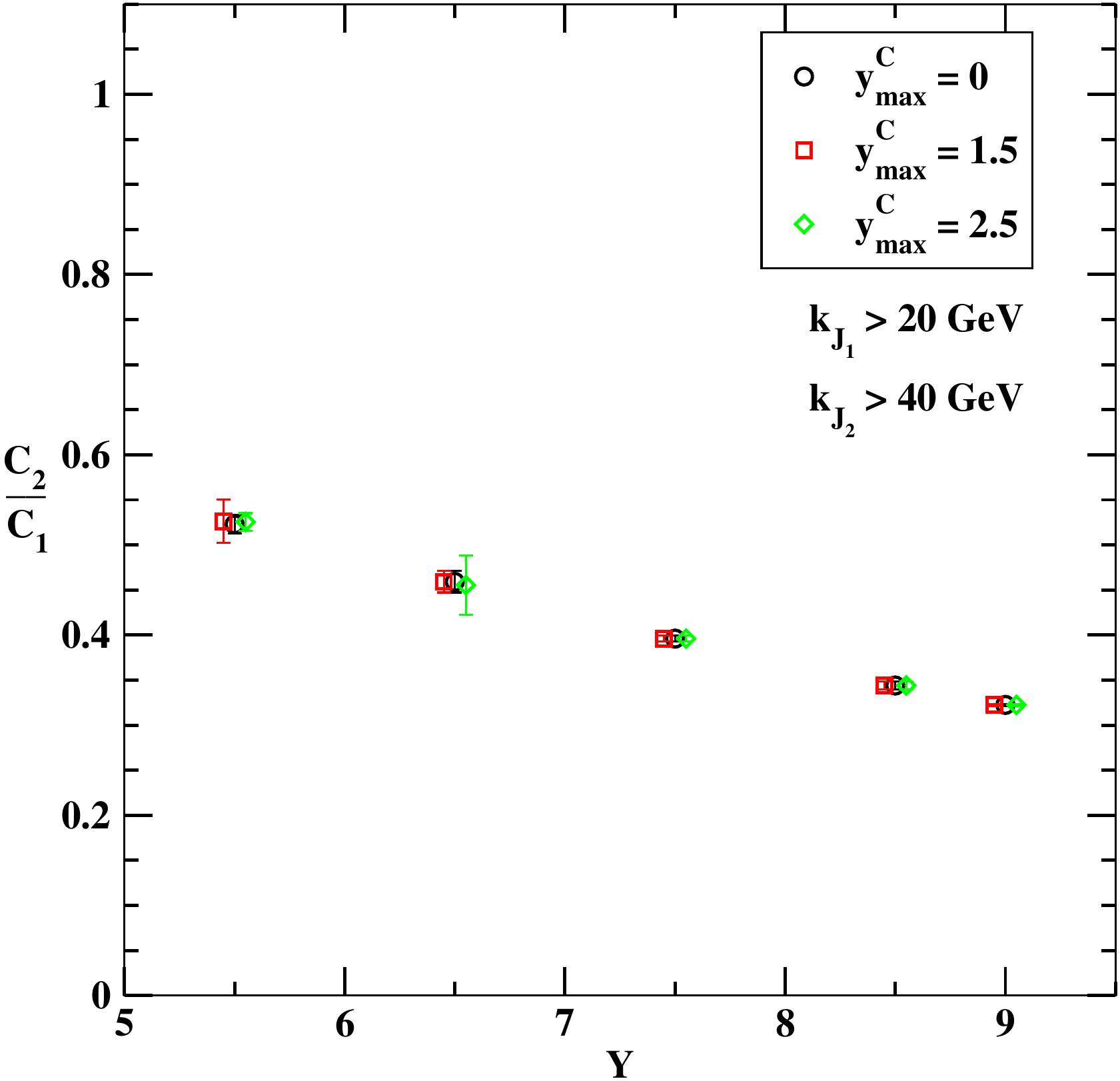}

   \includegraphics[scale=0.40]{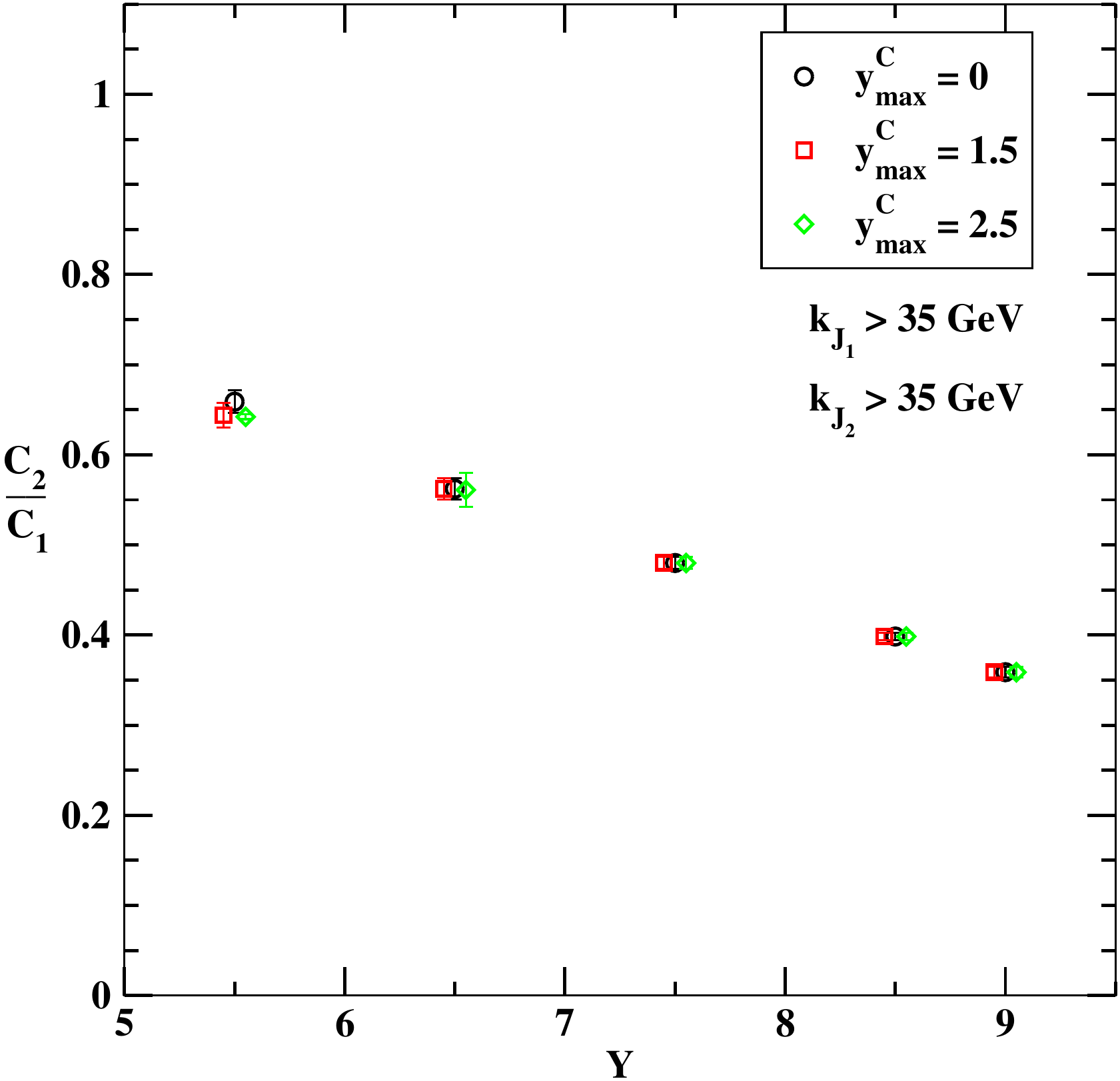}
\caption{$Y$-dependence of $C_2/C_1$ from the ``exact'' BLM method,
for all choices of the cuts on jet transverse momenta and of the central
rapidity region (data points have been slightly shifted along the horizontal
axis for the sake of readability; see Table~\ref{tab:C2C1_e}).}
\label{C2C1_e}
\end{figure}

\begin{figure}[p]
\centering

   \includegraphics[scale=0.40]{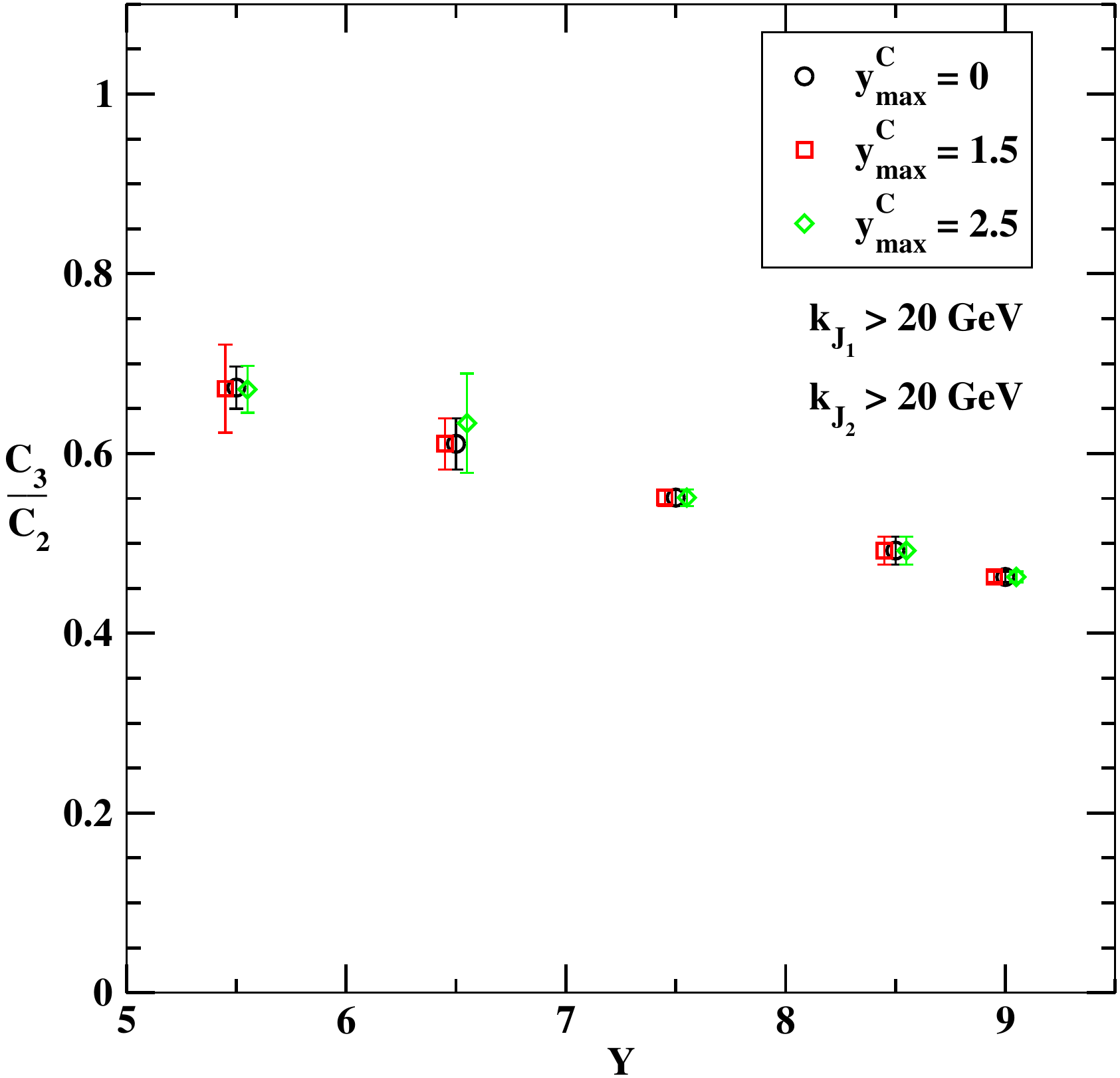}
   \includegraphics[scale=0.40]{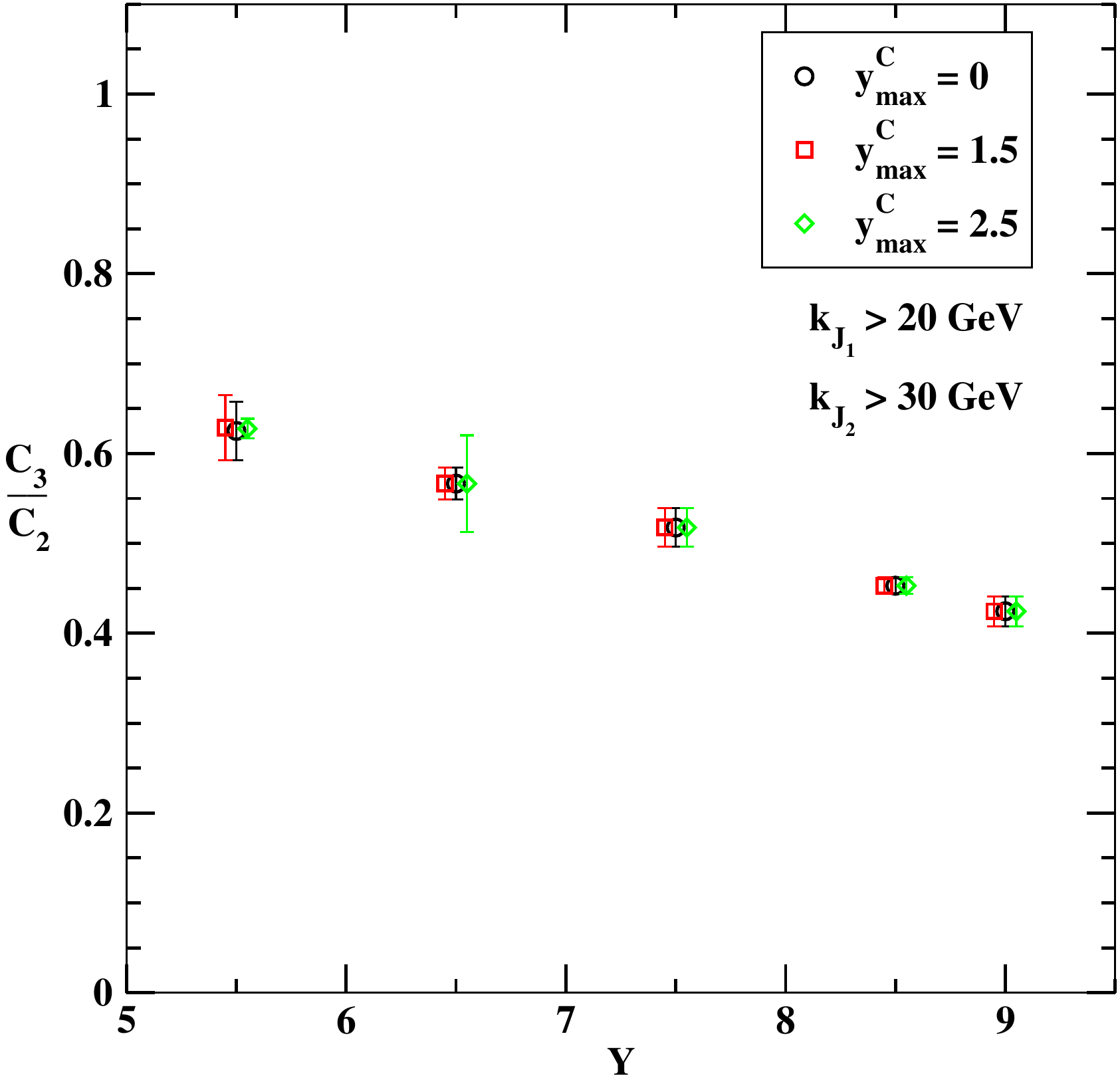}

   \includegraphics[scale=0.40]{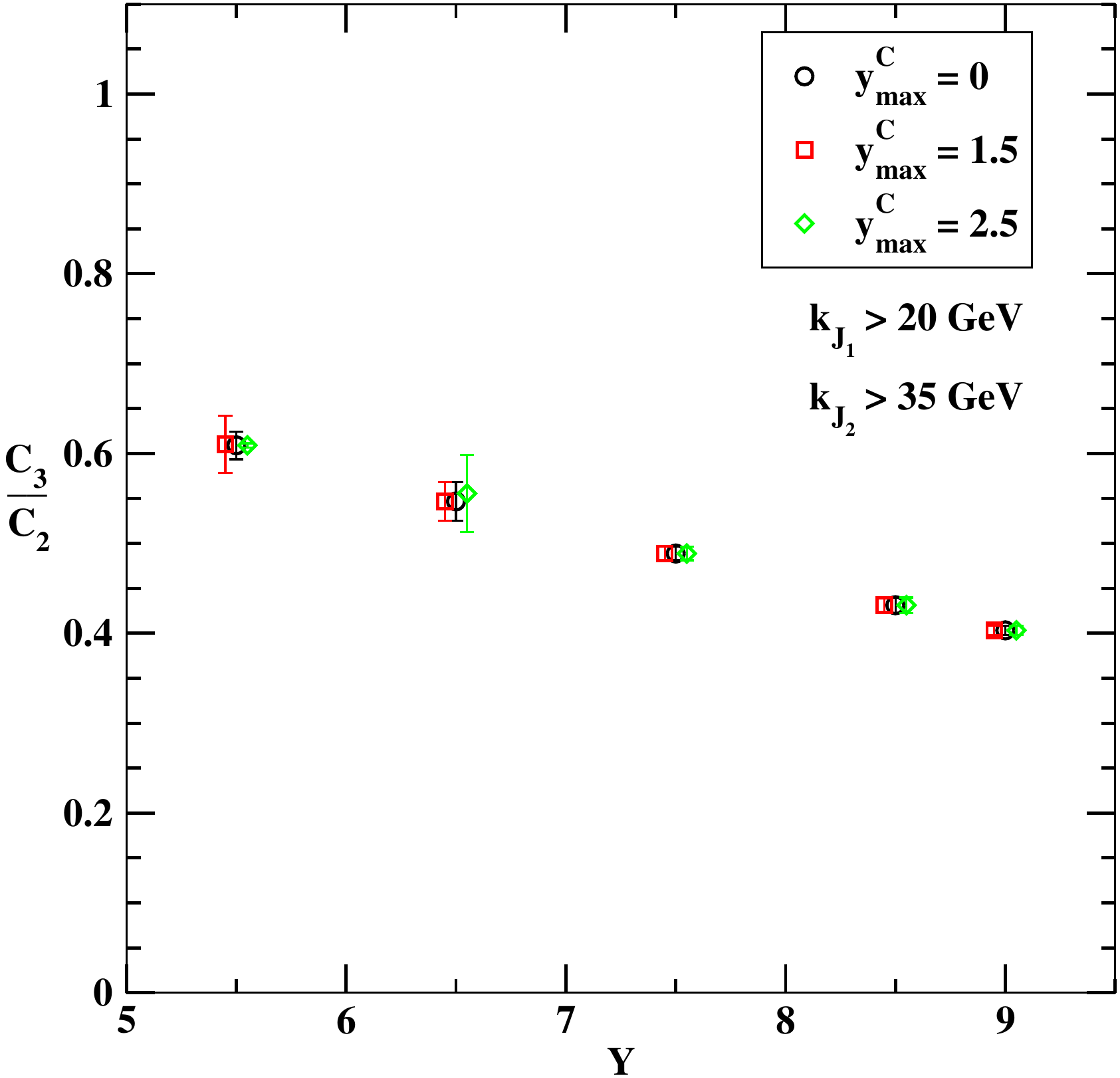}
   \includegraphics[scale=0.40]{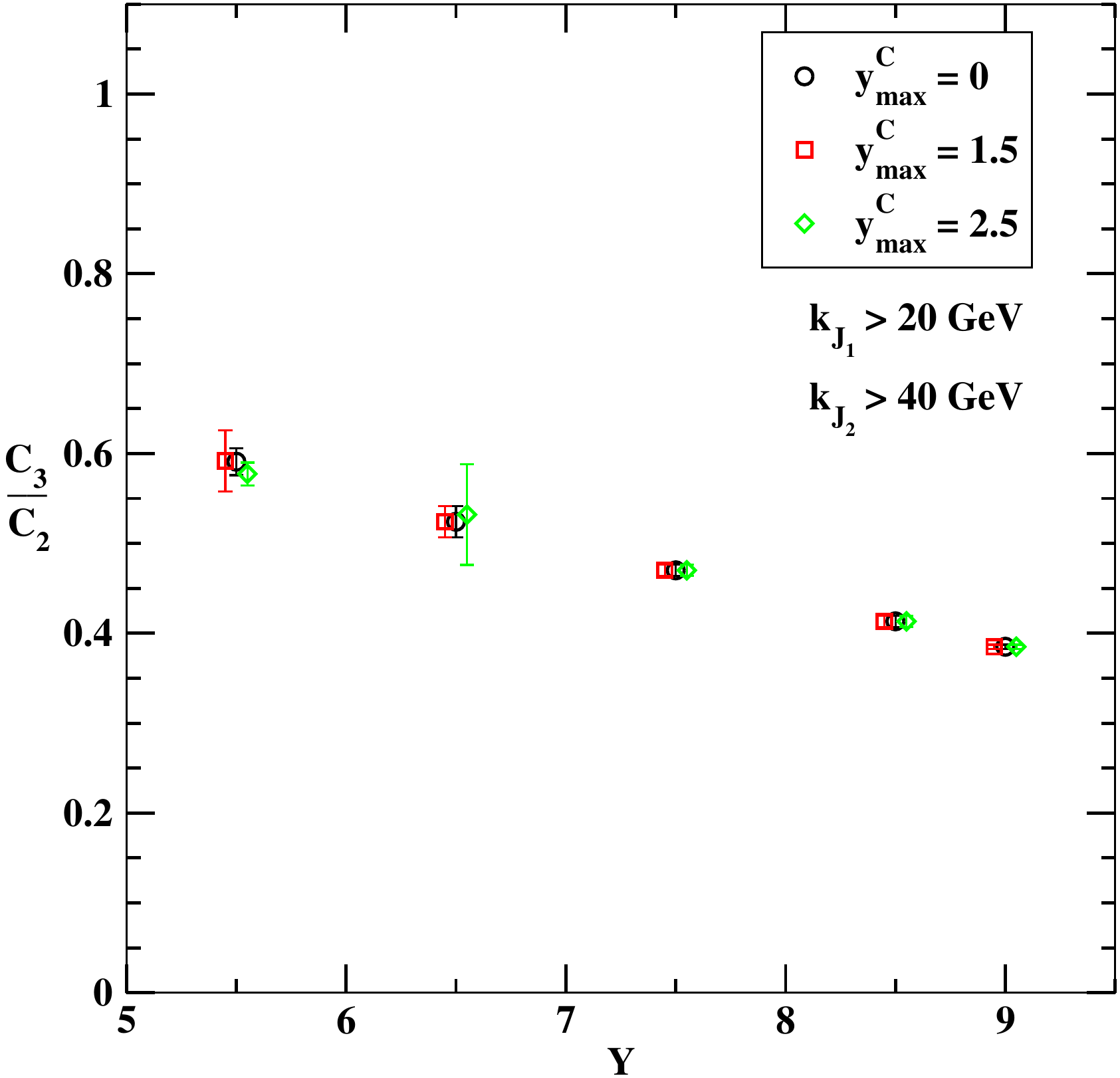}

   \includegraphics[scale=0.40]{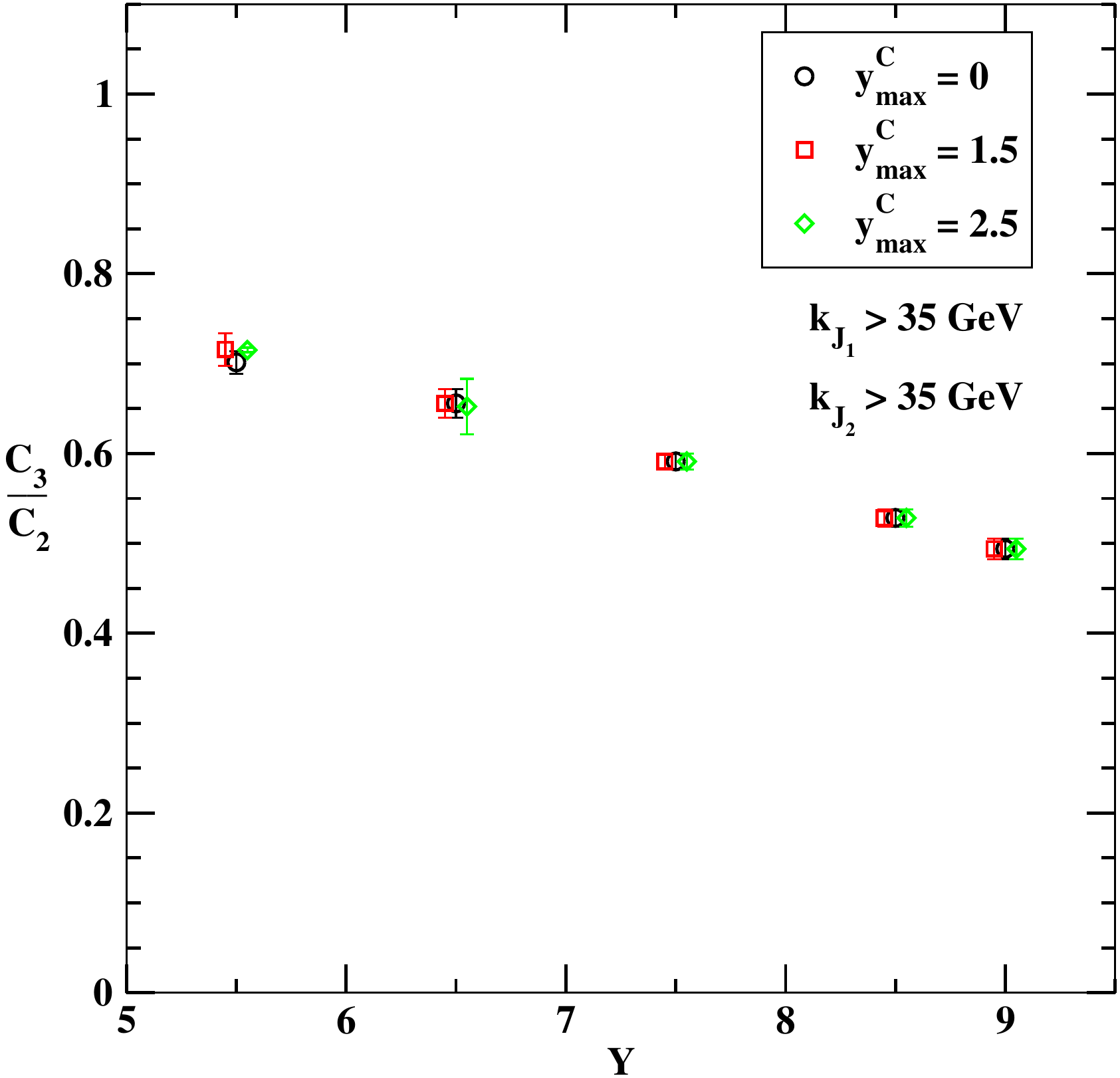}
\caption{$Y$-dependence of $C_3/C_2$ from the ``exact'' BLM method,
for all choices of the cuts on jet transverse momenta and of the central
rapidity region (data points have been slightly shifted along the horizontal
axis for the sake of readability; see Table~\ref{tab:C3C2_e}).}
\label{C3C2_e}
\end{figure}

We summarize our results in Tables~\ref{tab:2020_2.5}-\ref{tab:C3C2_e}
and in Figs.~\ref{2020_2.5}-\ref{C3C2_e}. From Table~\ref{tab:2020_2.5}
(and Fig.~\ref{2020_2.5}) we can see that the different variants of 
implementation of the BLM method give predictions which deviate at the level 
of $\sim 10\%$ for $C_0$ and at the level of $\sim 5\%$ for $C_1/C_0$, while 
they basically agree within errors for all other ratios $R_{nm}$. For this 
reason, all remaining Tables (and Figures) refer to the ``exact'' BLM case only.
Table~\ref{tab:C0_e} (and Fig.~\ref{C0_e}) show, quite reasonably, that
for all choices of the cuts on jet transverse momenta, the larger is
$y^{\rm C}_{\rm max}$, the lower is the total cross section $C_0$, up the
value of $Y$ is reached where the presence of cut of the central rapidity
region becomes ineffective. All remaining Tables (and Figures) unanimously
show that all ratios $R_{nm}$ remain unaffected by the cut on the central
rapidity region, over the entire region of values of $Y$. This is obvious
for the values of $Y$ large enough to be insensitive to the very presence of
a non-zero $y^{\rm C}_{\rm max}$, but it is unexpectedly true also for the
lower values of $Y$. 

The latter point means that in our approach, {\it i.e.} NLA BFKL with BLM 
optimization, the cut on jet central rapidities leads to a proportional 
reduction of both the total cross section, $C_0$, and the other coefficients 
$C_1, C_2, C_3$, which parametrize the azimuthal angle distribution. In other 
words in our approach, the central cut only reduces the value of the total 
cross section, but does not affect the azimuthal angle distribution of dijets. 
It would be very interesting to study whether such feature remains true also in 
other approaches, both within the BFKL approach, but using different ideas 
about the inclusion of the physics beyond NLA, and also in other, non-BFKL 
schemes, like fixed-order DGLAP or approaches using $k_T$-factorization for 
the central jet production.  

\section{Conclusions}
\label{conclusions}

In this paper we have considered the Mueller-Navelet jet production
process at LHC at the center-of-mass energy of 13~TeV and have produced
predictions for total cross sections and several azimuthal correlations and 
ratios between them in full NLA BFKL approach, in a theoretical setup in which 
jet vertices where taken in the so-called ``small-cone approximation'' and the 
BFKL series was optimized adopting the BLM method to fix, at a common value,
the renormalization and the factorization scales.

It is known that BFKL predictions for the Mueller-Navelet process suffer from 
large uncertainties due to basically our disability to resum BFKL energy 
logarithms beyond NLA in a model-independent way.
In this situation one needs to rely on some approaches to optimization of 
perturbative series.
Here we have used the BLM method which was previously quite successful in 
describing the LHC 7~TeV data on jet angular correlations. We hope that 
the forthcoming LHC analysis at 13~TeV will shed a new light on the issue and 
will allow to better discriminate among theoretical ideas about the
BFKL physics beyond the NLA approximation. In this respect we believe that it 
could be advantageous if the comparison of theory predictions with the data 
would be done in a kinematic range where theoretical calculations do 
not have other uncertainties except the ones mentioned above.    

Therefore here, differently from all previous studies of the same kind, we 
considered in our analysis the effect of excluding the possibility that one 
of the two detected jets be produced in the central rapidity region. 
Central jets originate from small-$x$ partons, and the collinear approach for 
the description of the Mueller-Navelet jet vertices may be not good at 
small $x$. 
The outcome of our analysis is that, for two reasonable ways to define the 
extension of the central region:
a) the total cross section, $C_0$, is strongly reduced by the ``exclusion cuts''
in the range ($Y<5.5$) where they are effective;
b) on the other hand, in the same kinematics, the difference with respect to 
the case of no central rapidity exclusion is invisible in azimuthal 
correlations and in ratios between them.

We believe that it would be very interesting to confront these conclusions 
with LHC data.   

\vspace{1.0cm} \noindent
{\Large \bf Acknowledgments} \vspace{0.5cm}

We thank G.~Safronov for fruitful discussions.
\\
\indent
The work of D.I. was supported in part by the grant RFBR-15-02-05868-a.



%%%%%%%%%%%%%%%%%%%%%%%%%%%%%%%% TABLES %%%%%%%%%%%%%%%%%%%%%%%%%%%%%%%%%%

%%%%%%%% T 20-20 2.5 %%%%%%%%
\begin{table}[p]
\centering
\caption{$C_0$ [nb] and ratios $C_n/C_m$ for $k_{J_1,\rm min}=k_{J_2,\rm min}=20$~GeV 
and $y^{\rm C}_{\rm max}=2.5$, for the three variants of the BLM method (see
Fig.~\ref{2020_2.5}).}
\label{tab:2020_2.5}
\begin{tabular}{c|c|lll}
%\hline\noalign{\smallskip}
\toprule
          & $Y$ & BLM$_{a}$ & BLM$_{b}$ & BLM$_{\rm exact}$ \\
\midrule
          & 5.5 & 1353.2(5.6) & 1413.2(3.2) & 1318(16)    \\
          & 6.5 & 1778(23)    & 1877(13)    & 1720(49)    \\
$C_0$     & 7.5 & 834.6(2.8)  & 893.7(2.0)   & 803.4(6.6)  \\
          & 8.5 & 140.06(25)  & 152.03(18)  & 133.91(78)  \\
          & 9.0 & 32.97(10)   & 36.16(12)   & 31.46(20)   \\
\midrule
          & 5.5 & 0.7641(68)  & 0.7434(37)  & 0.775(19)   \\
          & 6.5 & 0.674(17)   & 0.6546(87)  & 0.686(37)   \\
$C_1/C_0$ & 7.5 & 0.6005(44)  & 0.5775(22)  & 0.6104(99)  \\
          & 8.5 & 0.5339(19)  & 0.5092(11)  & 0.5422(64)  \\
          & 9.0 & 0.5091(27)  & 0.4823(23)  & 0.5174(65)  \\
\midrule
          & 5.5 & 0.4371(52)  & 0.4315(29)  & 0.450(18)   \\
          & 6.5 & 0.336(11)   & 0.3357(53)  & 0.3329(19)  \\
$C_2/C_0$ & 7.5 & 0.2638(27)  & 0.2625(13)  & 0.2611(35)  \\
          & 8.5 & 0.2052(11)  & 0.20452(59) & 0.1939(49)  \\
          & 9.0 & 0.1835(14)  & 0.1827(11)  & 0.1674(14)  \\
\midrule
          & 5.5 & 0.2761(45)  & 0.2691(26)  & 0.3019(68)  \\
          & 6.5 & 0.1934(74)  & 0.1907(37)  & 0.210(18)   \\
$C_3/C_0$ & 7.5 & 0.1383(20)  & 0.13708(80) & 0.144(29)   \\
          & 8.5 & 0.09796(70) & 0.09765(31) & 0.095(17)   \\
          & 9.0 & 0.08378(90) & 0.08361(63) & 0.0775(13)  \\
\midrule
          & 5.5 & 0.5721(71)  & 0.5804(42)  & 0.580(24)   \\
          & 6.5 & 0.499(15)   & 0.5128(76)  & 0.484(27)   \\
$C_2/C_1$ & 7.5 & 0.4393(47)  & 0.4546(19)  & 0.4278(55)  \\
          & 8.5 & 0.3844(21)  & 0.4017(11)  & 0.3576(91)  \\
          & 9.0 & 0.3605(23)  & 0.3788(19)  & 0.3236(27)  \\
\midrule
          & 5.5 & 0.632(13)   & 0.6236(74)  & 0.671(26)   \\
          & 6.5 & 0.575(25)   & 0.568(12)   & 0.634(55)   \\
$C_3/C_2$ & 7.5 & 0.5241(93)  & 0.5221(32)  & 0.5509(92)  \\
          & 8.5 & 0.4773(41)  & 0.4775(18)  & 0.492(16)   \\
          & 9.0 & 0.4565(55)  & 0.4577(29)  & 0.4627(59)  \\
\bottomrule
\end{tabular}
\end{table}

%%%%%%%% T C0_exact %%%%%%%%
\begin{table}[p]
\centering
\caption{Values of $C_0$ [nb] from the ``exact'' BLM method, for all
choices of the cuts on jet transverse momenta and of the central rapidity region
(see Fig.~\ref{C0_e}).}
\label{tab:C0_e}
\begin{tabular}{c|c|c|lll}
%\hline\noalign{\smallskip}
\toprule
$k_{J_1,\rm min}$ & $k_{J_2,\rm min}$ & $Y$  & $y^{\rm C}_{\rm max}=0$
               & $y^{\rm C}_{\rm max}=1.5$ & $y^{\rm C}_{\rm max}=2.5$ \\
\midrule
       &        & 3.5 & 46100(950)  & 5498(110)   & -           \\
       &        & 4.5 & 20410(290)  & 8200(130)   & -           \\
       &        & 5.5 & 8270(130)   & 6120(110)   & 1318(16)    \\
20 GeV & 20 GeV & 6.5 & 2902(31)    & 2902(31)    & 1720(49)    \\
       &        & 7.5 & 803.4(6.6)  & 803.4(6.6)  & 803.4(6.6)  \\
       &        & 8.5 & 133.91(78)  & 133.91(78)  & 133.91(78)  \\
       &        & 9.0 & 31.46(20)   & 31.46(20)   & 31.46(20)   \\
\midrule
       &        & 3.5 & 15000(270)  & 1842(27)    & -           \\
       &        & 4.5 & 6734(73)    & 2779(33)    & -           \\
       &        & 5.5 & 2701(51)    & 2030(34)    & 442.3(3.4)  \\
20 GeV & 30 GeV & 6.5 & 919.8(9.2)  & 919.8(9.2)  & 555(13)     \\
       &        & 7.5 & 240.8(1.6)  & 240.8(1.6)  & 240.8(1.6)  \\
       &        & 8.5 & 36.44(13)   & 36.44(13)   & 36.44(13)   \\
       &        & 9.0 & 7.801(53)   & 7.801(53)   & 7.801(53)   \\
\midrule
       &        & 3.5 & 8090(160)   & 1050(20)    & -           \\
       &        & 4.5 & 3793(54)    & 1598(21)    & -           \\
       &        & 5.5 & 1534(26)    & 1169(16)    & 256.0(2.1)  \\
20 GeV & 35 GeV & 6.5 & 520.6(6.2)  & 520.6(6.2)  & 318.5(6.9)  \\
       &        & 7.5 & 134.2(1.1)  & 134.2(1.1)  & 134.2(1.1)  \\
       &        & 8.5 & 19.422(98)  & 19.422(98)  & 19.422(98)  \\
       &        & 9.0 & 3.9601(23)  & 3.9601(23)  & 3.9601(23)  \\
\midrule
       &        & 3.5 & 4627(86)    & 595.3(7.3)  & -           \\
       &        & 4.5 & 2137(31)    & 912(10)     & -           \\
       &        & 5.5 & 872(13)     & 668(10)     & 146.68(94)  \\
20 GeV & 40 GeV & 6.5 & 295.4(2.7)  & 295.4(2.7)  & 181.6(4.1)  \\
       &        & 7.5 & 74.75(37)   & 74.75(37)   & 74.75(37)   \\
       &        & 8.5 & 10.362(30)  & 10.362(30)  & 10.362(30)  \\
       &        & 9.0 & 1.9980(45)  & 1.9980(45)  & 1.9980(45)  \\
\midrule
       &        & 3.5 & 4286(36)    & 544.7(6.0)  & -           \\
       &        & 4.5 & 1618(13)    & 690.9(3.3)  & -           \\
       &        & 5.5 & 555.2(4.1)  & 429.0(3.6)  & 94.48(13)   \\
35 GeV & 35 GeV & 6.5 & 161.8(1.2)  & 161.8(1.2)  & 101.5(1.1)  \\
       &        & 7.5 & 35.70(16)   & 35.70(16)   & 35.70(16)   \\
       &        & 8.5 & 4.2843(98)  & 4.2843(98)  & 4.2843(98)  \\
       &        & 9.0 & 0.7579(23)  & 0.7579(23)  & 0.7579(23)  \\
\bottomrule
\end{tabular}
\end{table}

%%%%%%%% T C1/C0_exact %%%%%%%%
\begin{table}[p]
\centering
\caption{Values of $C_1/C_0$ from the ``exact'' BLM method, for all
choices of the cuts on jet transverse momenta and of the central rapidity region
(see Fig.~\ref{C1C0_e}).}
\label{tab:C1C0_e}
\begin{tabular}{c|c|c|lll}
%\hline\noalign{\smallskip}
\toprule
$k_{J_1,\rm min}$ & $k_{J_2,\rm min}$ & $Y$  & $y^{\rm C}_{\rm max}=0$
               & $y^{\rm C}_{\rm max}=1.5$ & $y^{\rm C}_{\rm max}=2.5$ \\
\midrule
       &        & 3.5 & 0.988(37)   & 0.975(35)   & -           \\
       &        & 4.5 & 0.885(25)   & 0.874(27)   & -           \\
       &        & 5.5 & 0.785(25)   & 0.778(31)   & 0.775(19)   \\
20 GeV & 20 GeV & 6.5 & 0.692(18)   & 0.692(18)   & 0.686(37)   \\
       &        & 7.5 & 0.6104(99)  & 0.6104(99)  & 0.6104(99)  \\
       &        & 8.5 & 0.5423(64)  & 0.5423(64)  & 0.5423(64)  \\
       &        & 9.0 & 0.5174(64)  & 0.5174(64)  & 0.5174(64)  \\
\midrule
       &        & 3.5 & 1.004(31)   & 0.989(28)   & -           \\
       &        & 4.5 & 0.896(18)   & 0.886(20)   & -           \\
       &        & 5.5 & 0.799(27)   & 0.792(27)   & 0.783(10)   \\
20 GeV & 30 GeV & 6.5 & 0.710(13)   & 0.710(13)   & 0.702(33    \\
       &        & 7.5 & 0.6321(83)  & 0.6321(83)  & 0.6321(83)  \\
       &        & 8.5 & 0.5717(45)  & 0.5717(45)  & 0.5717(45)  \\
       &        & 9.0 & 0.5543(70)  & 0.5543(70)  & 0.5543(70)  \\
\midrule
       &        & 3.5 & 1.051(37)   & 1.005(33)   & -           \\
       &        & 4.5 & 0.907(24)   & 0.892(24)   & -           \\
       &        & 5.5 & 0.803(28)   & 0.795(22)   &  0.788(13)  \\
20 GeV & 35 GeV & 6.5 & 0.712(16)   & 0.712(16)   &  0.704(31)  \\
       &        & 7.5 & 0.636(10)   & 0.636(10)   & 0.636(10)   \\
       &        & 8.5 & 0.5803(56)  & 0.5803(56)  & 0.5803(56)  \\
       &        & 9.0 & 0.5679(74)  & 0.5679(74)  & 0.5679(74)  \\
\midrule
       &        & 3.5 & 1.043(35)   & 1.021(22)   & -           \\
       &        & 4.5 & 0.916(25)   & 0.899(20)   & -           \\
       &        & 5.5 & 0.808(22)   & 0.798(24)   & 0.791(10)   \\
20 GeV & 40 GeV & 6.5 & 0.714(12)   & 0.714(12)   & 0.705(31)   \\
       &        & 7.5 & 0.6383(64)  & 0.6383(64)  & 0.6383(64)  \\
       &        & 8.5 & 0.5875(35)  & 0.5875(35)  & 0.5875(35)  \\
       &        & 9.0 & 0.5804(25)  & 0.5804(25)  & 0.5804(25)  \\
\midrule
       &        & 3.5 & 0.963(16)   & 0.952(18)   & -           \\
       &        & 4.5 & 0.883(14)   & 0.8722(82)  & -           \\
       &        & 5.5 & 0.798(13)   & 0.792(12)   & 0.7866(22)  \\
35 GeV & 35 GeV & 6.5 & 0.718(11)   & 0.718(11)   &  0.709(16)  \\
       &        & 7.5 & 0.6478(53)  & 0.6478(53)  & 0.6478(53)  \\
       &        & 8.5 & 0.5972(26)  & 0.5972(26)  & 0.5972(26)  \\
       &        & 9.0 & 0.5886(33)  & 0.5886(33)  & 0.5886(33)  \\
\bottomrule
\end{tabular}
\end{table}

%%%%%%%% T C2/C0_exact %%%%%%%%
\begin{table}[p]
\centering
\caption{Values of $C_2/C_0$ from the ``exact'' BLM method, for all
choices of the cuts on jet transverse momenta and of the central rapidity region
(see Fig.~\ref{C2C0_e}).}
\label{tab:C2C0_e}
\begin{tabular}{c|c|c|lll}
%\hline\noalign{\smallskip}
\toprule
$k_{J_1,\rm min}$ & $k_{J_2,\rm min}$ & $Y$ & $y^{\rm C}_{\rm max}=0$
               & $y^{\rm C}_{\rm max}=1.5$ & $y^{\rm C}_{\rm max}=2.5$ \\
\midrule
       &        & 3.5 & 0.749(25)   & 0.730(30)   & -           \\
       &        & 4.5 & 0.594(23)   & 0.581(24)   & -           \\
       &        & 5.5 & 0.458(13)   & 0.454(27)   & 0.450(18)   \\
20 GeV & 20 GeV & 6.5 & 0.350(13)   & 0.350(13)   & 0.332(19)   \\
       &        & 7.5 & 0.2611(35)  & 0.2611(35)  & 0.2611(35)  \\
       &        & 8.5 & 0.1939(49)  & 0.1939(49)  & 0.1939(49)  \\
       &        & 9.0 & 0.1674(14)  & 0.1674(14)  & 0.1674(14)  \\
\midrule
       &        & 3.5 & 0.727(27)   & 0.719(26)   & -           \\
       &        & 4.5 & 0.575(15)   & 0.565(17)   & -           \\
       &        & 5.5 & 0.450(20)   & 0.443(21)   & 0.4398(98)  \\
20 GeV & 30 GeV & 6.5 & 0.3483(94)  & 0.3483(94)  & 0.343(24)   \\
       &        & 7.5 & 0.2683(53)  & 0.2683(53)  & 0.2683(53)  \\
       &        & 8.5 & 0.2083(30)  & 0.2083(30)  & 0.2083(30)  \\
       &        & 9.0 & 0.1872(39)  & 0.1872(39)  & 0.1872(39)  \\
\midrule
       &        & 3.5 & 0.750(22)   & 0.714(29)   & -           \\
       &        & 4.5 & 0.563(20)   & 0.555(20)   & -           \\
       &        & 5.5 & 0.435(11)   & 0.430(17)   & 0.4268(40)  \\
20 GeV & 35 GeV & 6.5 & 0.337(12)   & 0.337(12)   & 0.331(20)   \\
       &        & 7.5 & 0.2602(32)  & 0.2602(32)  & 0.2602(32)  \\
       &        & 8.5 & 0.2059(37)  & 0.2059(37)  & 0.2059(37)  \\
       &        & 9.0 & 0.1874(15)  & 0.1874(15)  & 0.1874(15)  \\
\midrule
       &        & 3.5 & 0.727(21)   & 0.710(19)   & -           \\
       &        & 4.5 & 0.560(17)   & 0.546(16)   & -           \\
       &        & 5.5 & 0.4225(99)  & 0.420(20)   & 0.4158(75)  \\
20 GeV & 40 GeV & 6.5 & 0.3276(91)  & 0.3276(91)  & 0.321(23)   \\
       &        & 7.5 & 0.2528(22)  & 0.2528(22)  & 0.2528(22)  \\
       &        & 8.5 & 0.2021(26)  & 0.2021(26)  & 0.2021(26)  \\
       &        & 9.0 & 0.18712(7)  & 0.18712(7)  & 0.18712(7)  \\
\midrule
       &        & 3.5 & 0.778(16)   & 0.766(16)   & -           \\
       &        & 4.5 & 0.642(12)   & 0.6321(85)  & -           \\
       &        & 5.5 & 0.5260(94)  & 0.510(12)   & 0.5051(20)  \\
35 GeV & 35 GeV & 6.5 & 0.4038(86)  & 0.4038(86)  & 0.398(13)   \\
       &        & 7.5 & 0.3109(45)  & 0.3109(45)  & 0.3109(45)  \\
       &        & 8.5 & 0.2379(25)  & 0.2379(25)  & 0.2379(25)  \\
       &        & 9.0 & 0.2112(37)  & 0.2112(37)  & 0.2112(37)  \\
\bottomrule
\end{tabular}
\end{table}

%%%%%%%% T C3/C0_exact %%%%%%%%
\begin{table}[p]
\centering
\caption{Values of $C_3/C_0$ from the ``exact'' BLM method, for all
choices of the cuts on jet transverse momenta and of the central rapidity region
(see Fig.~\ref{C3C0_e}).}
\label{tab:C3C0_e}
\begin{tabular}{c|c|c|lll}
%\hline\noalign{\smallskip}
\toprule
$k_{J_1,\rm min}$ & $k_{J_2,\rm min}$ & $Y$ & $y^{\rm C}_{\rm max}=0$
               & $y^{\rm C}_{\rm max}=1.5$ & $y^{\rm C}_{\rm max}=2.5$ \\
\midrule
       &        & 3.5 & 0.593(22)   & 0.577(19)   & -           \\
       &        & 4.5 & 0.432(13)   & 0.425(14)   & -           \\
       &        & 5.5 & 0.308(12)   & 0.305(15)   & 0.3019(68)  \\
20 GeV & 20 GeV & 6.5 & 0.2139(67)  & 0.2139(67)  & 0.210(18)   \\
       &        & 7.5 & 0.1439(29)  & 0.1439(29)  & 0.1439(29)  \\
       &        & 8.5 & 0.0954(17)  & 0.0954(17)  & 0.0954(17)  \\
       &        & 9.0 & 0.0775(13)  & 0.0775(13)  & 0.0775(13)  \\
\midrule
       &        & 3.5 & 0.551(26)   & 0.544(14)   & -           \\
       &        & 4.5 & 0.3950(88)  & 0.3896(97)  & -           \\
       &        & 5.5 & 0.281(13)   & 0.278(12)   & 0.276(3)    \\
20 GeV & 30 GeV & 6.5 & 0.1973(48)  & 0.1973(48)  & 0.194(14)   \\
       &        & 7.5 & 0.1389(49)  & 0.1389(49)  & 0.1389(49)  \\
       &        & 8.5 & 0.0944(13)  & 0.0944(13)  & 0.0944(13)  \\
       &        & 9.0 & 0.0795(25)  & 0.0795(25)  & 0.0795(25)  \\
\midrule
       &        & 3.5 & 0.555(19)   & 0.528(15)   & -           \\
       &        & 4.5 & 0.377(11)   & 0.3724(94)  & -           \\
       &        & 5.5 & 0.2652(90)  & 0.263(10)   & 0.2599(30)  \\
20 GeV & 35 GeV & 6.5 & 0.1842(48)  & 0.1842(48)  & 0.184(11)   \\
       &        & 7.5 & 0.1272(24)  & 0.1272(24)  & 0.1272(24)  \\
       &        & 8.5 & 0.0888(11)  & 0.0888(11)  & 0.0888(11)  \\
       &        & 9.0 & 0.0756(12)  & 0.0756(12)  & 0.0756(12)  \\
\midrule
       &        & 3.5 & 0.529(18)   & 0.520(21)   & -           \\
       &        & 4.5 & 0.364(10)   & 0.3585(79)  & -           \\
       &        & 5.5 & 0.2496(80)  & 0.249(11)   & 0.2400(40)  \\
20 GeV & 40 GeV & 6.5 & 0.1717(41)  & 0.1717(41)  & 0.171(13)   \\
       &        & 7.5 & 0.1188(18)  & 0.1188(18)  & 0.1188(18)  \\
       &        & 8.5 & 0.0836(66)  & 0.0836(66)  & 0.0836(66)  \\
       &        & 9.0 & 0.0720(52)  & 0.0720(52)  & 0.0720(52)  \\
\midrule
       &        & 3.5 & 0.6478(76)  & 0.6360(95)  & -           \\
       &        & 4.5 & 0.4983(75)  & 0.4887(40)  & -           \\
       &        & 5.5 & 0.3690(55)  & 0.3652(69)  & 0.3613(84)  \\
35 GeV & 35 GeV & 6.5 & 0.2648(47)  & 0.2648(47)  & 0.2596(93)  \\
       &        & 7.5 & 0.1838(17)  & 0.1838(17)  & 0.1838(17)  \\
       &        & 8.5 & 0.1257(16)  & 0.1257(16)  & 0.1257(16)  \\
       &        & 9.0 & 0.1043(12)  & 0.1043(12)  & 0.1043(12)  \\
\bottomrule
\end{tabular}
\end{table}

%%%%%%%% T C2/C1_exact %%%%%%%%
\begin{table}[p]
\centering
\caption{Values of $C_2/C_1$ from the ``exact'' BLM method, for all
choices of the cuts on jet transverse momenta and of the central rapidity region
(see Fig.~\ref{C2C1_e}).}
\label{tab:C2C1_e}
\begin{tabular}{c|c|c|lll}
%\hline\noalign{\smallskip}
\toprule
$k_{J_1,\rm min}$ & $k_{J_2,\rm min}$ & $Y$ & $y^{\rm C}_{\rm max}=0$
               & $y^{\rm C}_{\rm max}=1.5$ & $y^{\rm C}_{\rm max}=2.5$ \\
\midrule
       &        & 3.5 & 0.759(21)   & 0.749(28)   & -           \\
       &        & 4.5 & 0.671(26)   & 0.665(28)   & -           \\
       &        & 5.5 & 0.583(17)   & 0.583(37)   & 0.580(24)   \\
20 GeV & 20 GeV & 6.5 & 0.506(21)   & 0.506(21)   & 0.484(27)   \\
       &        & 7.5 & 0.4278(55)  & 0.4278(55)  & 0.4278(55)  \\
       &        & 8.5 & 0.3576(91)  & 0.3576(91)  & 0.3576(91)  \\
       &        & 9.0 & 0.3236(27)  & 0.3236(27)  & 0.3236(27)  \\
\midrule
       &        & 3.5 & 0.724(23)   & 0.727(25)   & -           \\
       &        & 4.5 & 0.642(15)   & 0.638(18)   & -           \\
       &        & 5.5 & 0.563(23)   & 0.559(27)   & 0.561(11)   \\
20 GeV & 30 GeV & 6.5 & 0.491(13)   & 0.491(13)   & 0.489(34)   \\
       &        & 7.5 & 0.4245(83)  & 0.4245(83)  & 0.4245(83)  \\
       &        & 8.5 & 0.3644(54)  & 0.3644(54)  & 0.3644(54)  \\
       &        & 9.0 & 0.3377(67)  & 0.3377(67)  & 0.3377(67)  \\
\midrule
       &        & 3.5 & 0.713(19)   & 0.710(24)   & -           \\
       &        & 4.5 & 0.622(21)   & 0.623(22)   & -           \\
       &        & 5.5 & 0.542(14)   & 0.542(22)   & 0.5414(50)  \\
20 GeV & 35 GeV & 6.5 & 0.473(17)   & 0.473(17)   & 0.470(29)   \\
       &        & 7.5 & 0.4095(50)  & 0.4095(50)  & 0.4095(50)  \\
       &        & 8.5 & 0.3548(63)  & 0.3548(63)  & 0.3548(63)  \\
       &        & 9.0 & 0.3299(31)  & 0.3299(31)  & 0.3299(31)  \\
\midrule
       &        & 3.5 & 0.697(18)   & 0.695(16)   & -           \\
       &        & 4.5 & 0.612(17)   & 0.607(17)   & -           \\
       &        & 5.5 & 0.523(10)   & 0.526(24)   & 0.5256(96)  \\
20 GeV & 40 GeV & 6.5 & 0.459(12)   & 0.459(12)   & 0.455(33)   \\
       &        & 7.5 & 0.3960(33)  & 0.3960(33)  & 0.3960(33)  \\
       &        & 8.5 & 0.3441(45)  & 0.3441(45)  & 0.3441(45)  \\
       &        & 9.0 & 0.3224(12)  & 0.3224(12)  & 0.3224(12)  \\
\midrule
       &        & 3.5 & 0.809(16)   & 0.805(14)   & -           \\
       &        & 4.5 & 0.728(14)   & 0.7247(98)  & -           \\
       &        & 5.5 & 0.659(13)   & 0.644(14)   & 0.6421(27)  \\
35 GeV & 35 GeV & 6.5 & 0.562(12)   & 0.563(12)   & 0.561(19)   \\
       &        & 7.5 & 0.4799(67)  & 0.4799(67)  & 0.4799(67)  \\
       &        & 8.5 & 0.3984(40)  & 0.3984(40)  & 0.3984(40)  \\
       &        & 9.0 & 0.3588(61)  & 0.3588(61)  & 0.3588(61)  \\
\bottomrule
\end{tabular}
\end{table}

%%%%%%%% T C3/C2_exact %%%%%%%%
\begin{table}[p]
\centering
\caption{Values of $C_3/C_2$ from the ``exact'' BLM method, for all
choices of the cuts on jet transverse momenta and of the central rapidity region
(see Fig.~\ref{C3C2_e}).}
\label{tab:C3C2_e}
\begin{tabular}{c|c|c|lll}
%\hline\noalign{\smallskip}
\toprule
$k_{J_1,\rm min}$ & $k_{J_2,\rm min}$ & $Y$ & $y^{\rm C}_{\rm max}=0$
               & $y^{\rm C}_{\rm max}=1.5$ & $y^{\rm C}_{\rm max}=2.5$ \\
\midrule
       &        & 3.5 & 0.792(22)   & 0.790(26)   & -           \\
       &        & 4.5 & 0.727(30)   & 0.731(32)   & -           \\
       &        & 5.5 & 0.673(24)   & 0.672(49)   & 0.671(26)   \\
20 GeV & 20 GeV & 6.5 & 0.611(28)   & 0.611(28)   & 0.634(55)   \\
       &        & 7.5 & 0.5509(92)  & 0.5509(92)  & 0.5509(92)  \\
       &        & 8.5 & 0.492(16)   & 0.492(16)   & 0.492(16)   \\
       &        & 9.0 & 0.4627(59)  & 0.4627(59)  & 0.4627(59)  \\
\midrule
       &        & 3.5 & 0.758(37)   & 0.756(23)   & -           \\
       &        & 4.5 & 0.687(18)   & 0.689(22)   & -           \\
       &        & 5.5 & 0.625(32)   & 0.629(36)   & 0.628(11)   \\
20 GeV & 30 GeV & 6.5 & 0.566(18)   & 0.567(18)   & 0.566(54)   \\
       &        & 7.5 & 0.518(22)   & 0.518(22)   & 0.518(22)   \\
       &        & 8.5 & 0.4530(93)  & 0.4530(93)  & 0.4530(93)  \\
       &        & 9.0 & 0.424(17)   & 0.424(17)   & 0.424(17)   \\
\midrule
       &        & 3.5 & 0.741(19)   & 0.740(23)   & -           \\
       &        & 4.5 & 0.670(24)   & 0.671(23)   & -           \\
       &        & 5.5 & 0.609(15)   & 0.610(32)   & 0.6090(25)  \\
20 GeV & 35 GeV & 6.5 & 0.547(21)   & 0.547(21)   & 0.555(43)   \\
       &        & 7.5 & 0.4887(75)  & 0.4887(75)  & 0.4887(75)  \\
       &        & 8.5 & 0.4312(86)  & 0.4312(86)  & 0.4312(86)  \\
       &        & 9.0 & 0.4033(50)  & 0.4033(50)  & 0.4033(50)  \\
\midrule
       &        & 3.5 & 0.728(19)   & 0.732(32)   & -           \\
       &        & 4.5 & 0.650(18)   & 0.657(18)   & -           \\
       &        & 5.5 & 0.591(15)   & 0.592(34)   & 0.578(13)   \\
20 GeV & 40 GeV & 6.5 & 0.524(17)   & 0.524(17)   & 0.532(56)   \\
       &        & 7.5 & 0.4700(63)  & 0.4700(63)  & 0.4700(63)  \\
       &        & 8.5 & 0.4134(62)  & 0.4134(62)  & 0.4134(62)  \\
       &        & 9.0 & 0.3850(25)  & 0.3850(25)  & 0.3850(25)  \\
\midrule
       &        & 3.5 & 0.832(13)   & 0.830(11)   & -           \\
       &        & 4.5 & 0.776(14)   & 0.7731(94)  & -           \\
       &        & 5.5 & 0.701(13)   & 0.716(18)   & 0.7152(27)  \\
35 GeV & 35 GeV & 6.5 & 0.656(16)   & 0.656(16)   & 0.652(31)   \\
       &        & 7.5 & 0.5912(88)  & 0.5912(88)  & 0.5912(88)  \\
       &        & 8.5 & 0.5284(96)  & 0.5284(96)  & 0.5284(96)  \\
       &        & 9.0 & 0.4939(11)  & 0.4939(11)  & 0.4939(11)  \\
\bottomrule
\end{tabular}
\end{table}

%%%%%%%%%%%%%%%%%%%%%% FIGURES %%%%%%%%%%%%%%%%%%%%%%%%%%%%%%%%%%%%%%%%%%

%%%%%%%% F 20-20 2.5 %%%%%%%%

%%%%%%%% F C0 exact %%%%%%%%

%%%%%%%% F C1/C0 exact %%%%%%%%

%%%%%%%% F C2/C0 exact %%%%%%%%

%%%%%%%% F C3/C0 exact %%%%%%%%

%%%%%%%% F C2/C1 exact %%%%%%%%

%%%%%%%% F C3/C2 exact %%%%%%%%

\end{document}